\documentclass[aps,twocolumn,groupedaddress]{revtex4-2}

\usepackage{amsmath}
\usepackage{amssymb}
\usepackage{physics}
\usepackage{caption}
\usepackage{graphicx}
\usepackage[colorlinks = true, linkcolor = blue, urlcolor  = blue,
citecolor = blue, anchorcolor = blue]{hyperref}
\usepackage[caption=false]{subfig}
\usepackage{comment}
\usepackage{multirow}
\usepackage[font=small,labelfont=bf,justification=justified,format=plain]{caption}
\usepackage{xcolor}
\usepackage{tabularx}
\usepackage{ragged2e}
\usepackage{soul}

\begin{document}


\title{Shareability of Quantum Correlations in a Many-Body Spin System
with Two- and Three-Body Interactions}



\author{P. Kiran}
\thanks{}
\affiliation{Department of Physics, Indian Institute of Technology Dharwad, Dharwad, Karnataka, India - 580007}

\author{Harsha Miriam Reji}
\thanks{}
\affiliation{Department of Physics, Indian Institute of Technology Dharwad, Dharwad, Karnataka, India - 580007}

\author{Hemant S. Hegde}
\thanks{}
\affiliation{Department of Physics, Indian Institute of Technology Dharwad, Dharwad, Karnataka, India - 580007}

\author{R. Prabhu}
\thanks{}
\affiliation{Department of Physics, Indian Institute of Technology Dharwad, Dharwad, Karnataka, India - 580007}



\begin{abstract}
    The shareability of quantum correlations among the constituent
    parties of a multiparty quantum system is restricted by the quantum
    information theoretic concept called {\em monogamy}. Depending on
    the multiparty quantum systems, different measures of quantum
    correlations show disparate signatures for monogamy. We characterize
    the shareability of quantum correlations, from both
    entanglement-separability and information-theoretic kinds, in a
    multiparty quantum spin system containing two- and three-body
    interactions with respect to its system parameters and external
    applied magnetic field. Monogamy score in this system exhibits both
    monogamous and non-monogamous traits depending on the quantum
    correlation measure, strengths of system parameters and external
    magnetic field. The percentage of non-monogamous states when the
    information-theoretic quantum correlations are considered is higher
    than that of the entanglement-separability kind in allowed ranges of
    these variables. The integral powers of the quantum correlation
    measures for which the non-monogamous states become monogamous are
    identified.
\end{abstract}


\maketitle


\section{Introduction}
Quantum correlation~\cite{modi2012classical, horodecki2009quantum} is an
unassailable non-classical trait that is essential for the advancements
in quantum technologies~\cite{nielsen2000quantum}. Quantum correlations
present in finite and infinite dimensional quantum systems have been
harnessed to develop a multitude of quantum information protocols, such
as quantum dense coding~\cite{bennett1992communication}, quantum
teleportation~\cite{bennett1993teleporting}, quantum
computing~\cite{raussendorf2001one, briegel2009measurement}, quantum
sensors~\cite{degen2017quantum}, etc. Many-body quantum systems, like
spin chains~\cite{sachdev2011quantum}, optical
lattices~\cite{lewenstein2007ultracold}, trapped
ions~\cite{leibfried2003quantum}, photons~\cite{ma2011quantum}, etc.,
have been exploited as a potential physical resource in realizing
several of these protocols. Quantum information tools have also been
helpful to uncover several properties present in many-body systems,
such as the detection of quantum phase
transition~\cite{sachdev2011quantum, amico2008entanglement,
rezakhani2010intrinsic, del2012assisted}, magnetic
properties~\cite{lemmens2003magnetic, manousakis1991spin}, and the
development of the numerical simulations using matrix product
states~\cite{klumper1991equivalence, klumper1992groundstate},
projected entangled pair states~\cite{verstraete2004valence-bond},
etc.

The quantification of quantum correlations present in quantum systems is
crucial for the theoretical understanding as well as for their practical
implications. The quantifying measures of quantum correlation can be
broadly classified into two types: the entanglement–separability and the
information-theoretic kinds. While
concurrence~\cite{hill1997entanglement,wootters1998entanglement},
distillable entanglement~\cite{rains1999rigorous}, logarithmic
negativity~\cite{zyczkowski1998volume, lee2000partial,
vidal2002computable, plenio2005logarithmic}, relative entropy of
entanglement~\cite{vedral1998entanglement, vedral1997quantifying,
vedral2002role} are a few examples of the former category, quantum
discord~\cite{henderson2001classical, ollivier2001quantum}, quantum work
deficit~\cite{oppenheim2002thermodynamical, horodecki2003local,
horodecki2005local, devetak2005distillation}, generalized entropic
measures~\cite{rossignoli2010generalized} such as R\'enyi entropy and
Tsallis entropy belong to the latter category. These two categories of
quantum correlation measures are very different in nature and show a
variety of mutually exclusive properties for a given quantum system.

In a multiparty quantum system, there can be restrictions on the sharing
of quantum correlations among its subsystems. For example, in a
tripartite system, $ABC$, if subsystem $A$ is identified as a node,
there can be a restriction on the sharing of quantum correlation between
this node and any other subsystem, i.e., between $A$ and $B$ or $A$ and
$C$. Such a restriction on shareability among the subsystems of a
multiparty system is analytically captured by the {\em monogamy} of
quantum correlations~\cite{koashi2004monogamy, dhar2017monogamy,
kim2012limitations}. However, such a node in a multiparty classical
system can share any amount of correlation with any other parties within
the same system. The monogamy of quantum correlation forms a potential
ingredient in various quantum information protocols, including quantum
cryptography~\cite{gisin2002quantum}, entanglement
distillation~\cite{bennett1996mixed}, quantum state and channel
discrimination~\cite{prabhu2012conditions, giorgi2011monogamy,
kumar2016conclusive}, and energy transfer in biological
processes~\cite{zhu2012multipartite, chanda2015time}. Furthermore, the
study of monogamy relation can be instrumental in capturing the
multipartite quantum correlation present in a many-body quantum spin
system~\cite{sadhukhan2016quantum, allegra2011quantum, xk2013monogamy,
qin2016renormalization, Batle2017}.  By simulating the ground state of the
transverse Ising spin system in a triangular configuration on a nuclear
magnetic resonance physical system, it has been experimentally
demonstrated that a monogamy-based multipartite quantum correlation
measure can distinguish between frustrated and non-frustrated spin
systems~\cite{rao2013multipartite}. The monogamy relation has also been
used to simulate and identify the quantum phase transitions present in
the anisotropic XXZ quantum spin model~\cite{xk2013monogamy}. 

There has been a steady progress in understanding and characterizing
various properties of many-body quantum
systems~\cite{sachdev2011quantum}.  Predominantly, quantum spin systems
with only two-body interactions are used to study quantum information
related aspects in such systems~\cite{wang2012quantum, rong2012quantum,
giorda2010gaussian, adesso2010quantum, bellomo2012dynamics}. There is a
new fledgling interest to study several quantum information related
aspects in quantum spin systems with both two- and three-body
interactions~\cite{titvinidze2003phase, pachos2004effective,
sende2010channel, liu2012chiral, dcruz2005chiral, tsomokos2008chiral,
peng2010ground-state, you2016quantum, lou2004quantum, lou2005quantum,
dealcantarabonfim2014quantum, derzhko2011exact, shi2009effects,
peng2009quantum, zhang2010entanglement}. Along with this, their
experimental realizations in cold polar
molecules~\cite{capogrosso2009phase} and optical
lattices~\cite{pachos2004three} has led to significant progress in
many-body quantum information. Such quantum spin systems with two- and
three-body interactions have the potential to be great physical
resources for realizing several networked quantum information protocols.

In this paper, we study the behaviour of various quantum correlation
measures and the shareability of quantum correlations among the
subsystems of a multiparty quantum spin system, which has both two- and
three-body interactions by varying the system parameters and an external
applied magnetic field. In Sec.~\ref{sec:hamiltonian}, a succinct
description of this multiparty quantum spin system is provided. In
Sec.~\ref{sec:monogamy}, the monogamy score for any quantum correlation
measure in three-party and multiparty setting is described. In
Sec.~\ref{sec:shareability}, the behaviour of quantum correlation
measures chosen from both entanglement-separability and
information-theoretic kinds and the shareability of these quantum
correlations in the multiparty system under consideration are discussed
in detail. We also infer the ranges of all system parameters in which
the system will be monogamous or non-monogamous for the chosen measures
of quantum correlations. In Sec.~\ref{sec:percentage}, the percentages
of non-monogamous states for various quantum correlation measures for
the entire range of system parameters are given. The integer power of
various quantum correlation measures, used in the $N$-party monogamy
equation defined for the states produced by varying the entire range of
system parameters, for which the monogamy relation changes from
non-monogamous to monogamous is also listed.  In Sec.~\ref{sec:finite}, the effect of finite system size on the transition points of monogamy scores of quantum correlation measures obtained for the system under consideration is provided. A conclusion is drawn in
Sec.~\ref{sec:conclusion} and the quantum correlation measures
considered here are defined in Appendix~\ref{sec:appendix}.

\section{The Model Hamiltonian}\label{sec:hamiltonian}

\begin{figure}
    \centering
    \includegraphics[scale=1]{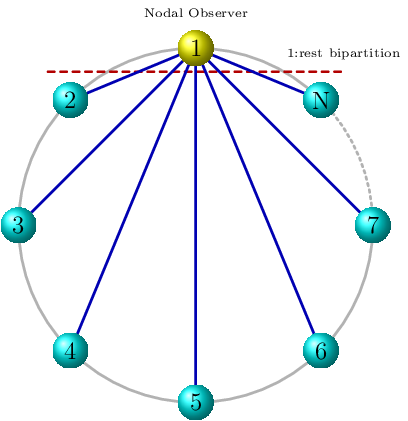}
    \captionsetup{justification=justified,singlelinecheck=off}
    \caption[]{(Color online) A schematic diagram for various possible
    bipartitions across which the quantum correlations are quantified in a
    cyclic one-dimensional multiparty quantum spin system with two- and
    three-body interactions. The quantum
    correlations in the 1:rest bipartition is captured by dashed (red)
    line, where spin 1 is chosen as a nodal observer. Each solid (blue)
    line represents the quantum correlations corresponding to the
    two-party reduced density matrices $\rho_{1k}$ ($k=2,3,...,N$).
    These quantum correlations, 
    $Q(\rho_{1:{\rm rest}})$ and $Q(\rho_{1k})$, appear in the
    definition of monogamy relation in Eq.~(\ref{eq:monogamy_N}).}
    \label{fig:monogamy}
\end{figure}

There is an increasing interest to study the many-body quantum spin systems with
two- and three-body interactions~\cite{titvinidze2003phase,
pachos2004effective, sende2010channel,  liu2012chiral,
dcruz2005chiral, tsomokos2008chiral, peng2010ground-state,
you2016quantum, lou2004quantum, lou2005quantum,
dealcantarabonfim2014quantum, derzhko2011exact, shi2009effects,
peng2009quantum, zhang2010entanglement, capogrosso2009phase,
pachos2004three}. Particularly, greater interest is attributed to the
spin systems which have three-body interactions of the type
XZX$+$YZY~\cite{titvinidze2003phase, derzhko2011exact} or
XZY$-$YZX~\cite{lou2004quantum, guo2011entanglement} along with two-body
interactions in them. It has also been demonstrated recently that
quantum communication can be enhanced if quantum spin systems with
XZY$-$YZX type of three-body interaction are
considered~\cite{xiang2010quantum}. 

The quantum spin system of interest in the current study contains
nearest-neighbour XX and YY type of two-body interactions and XZY$-$YZX
type of three-body interaction. We also consider the application of an
external magnetic field along the Z-direction of the system. The
Hamiltonian of this system is given by
\begin{eqnarray}
    H  &=& -\frac{J}{4} \sum_{n=1}^{N} \Big[  \sigma_{n}^{x}
    \sigma_{n+1}^{x} + \sigma_{n}^{y} \sigma_{n+1}^{y} \nonumber \\ & &
    + \frac{\alpha}{2} \qty(\sigma_{n-1}^{x} \sigma_{n}^{z}
    \sigma_{n+1}^{y} - \sigma_{n-1}^{y} \sigma_{n}^{z} \sigma_{n+1}^{x})
    \Big] -\frac{h}{2} \sum_{n=1}^{N} \sigma_{n}^{z}, \quad
\label{eq:ham}
\end{eqnarray}
where $J$ represents the two-body interaction strength, $\alpha$ being
the three-body interaction strength (scaled), $h$ is the external
applied magnetic field, $N$ is the total number of spins arranged in a
one-dimensional array, and $\sigma_j^a (a = x, y, z)$ are the Pauli spin
matrices at the site $j$. The periodic boundary condition is considered,
i.e., $\sigma_{N+1} = \sigma_1$.  We
numerically study the shareability of quantum correlations in this
system for a finite spin chain ($N=10$) by obtaining the ground state by
exact diagonalization.

	For the system defined in Eq.~(\ref{eq:ham}), the ground state and various physical quantities has been analytically
	obtained in Refs.~\cite{lou2004quantum, cheng2010disentanglement, yin2020quantum} when the system is of infinite chain length. This system is also known
	to exhibit quantum phase transitions~\cite{ding2014quantum, dai2020spin}
	and chiral phases~\cite{liu2012chiral, dcruz2005chiral}. This
	multiparty quantum spin system has been studied for its entanglement
	properties, like pairwise entanglement~\cite{guo2011pairwise} and
	thermal entanglement~\cite{fu2017effect}. The entanglement dynamics of the finite chain length of this system at finite temperature is studied in~\cite{guo2011entanglement}. Also, using the block-block entanglement, it has been shown that there exists a quantum phase transition at $\alpha=1$ and also the studies of its magnetization reveals the presence of four phases in this system~\cite{lou2006block}. It has been
	proposed in~\cite{bera2012multisite} that the genuine multipartite
	entanglement measure computed for this system can be an indicator of
	quantum phase transitions present in it.

\section{Monogamy Score of Quantum Correlations}\label{sec:monogamy}

\begin{figure*}[t]
\label{fig:ems}
  \subfloat[$h=0$]{\includegraphics[width=0.25\textwidth]{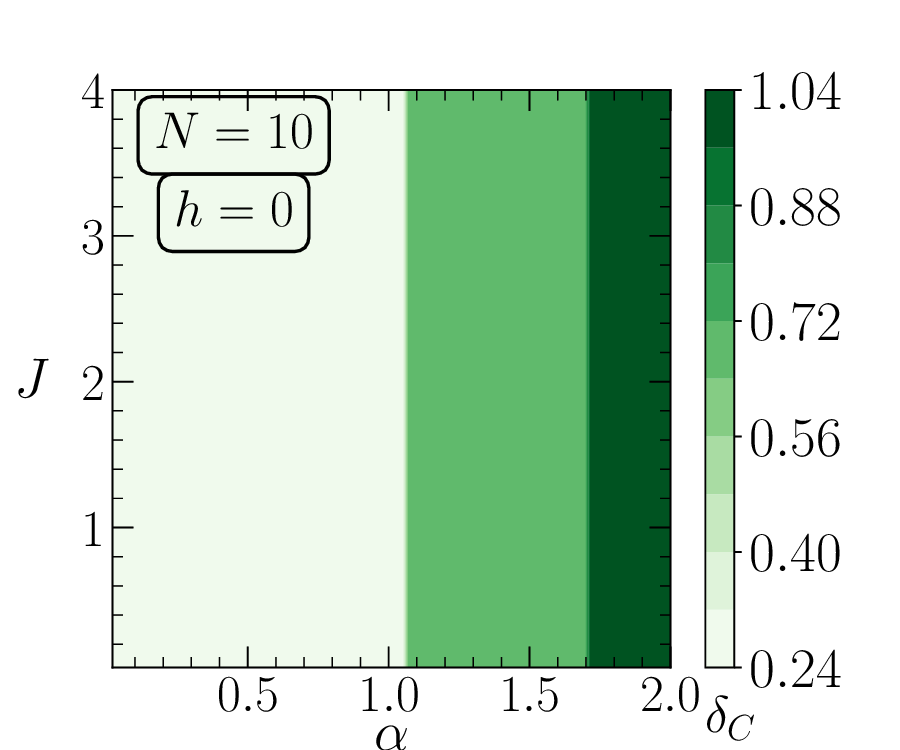}
    \label{fig:ems_10_0}}
  \subfloat[$h=0.2$]{\includegraphics[width=0.25\textwidth]{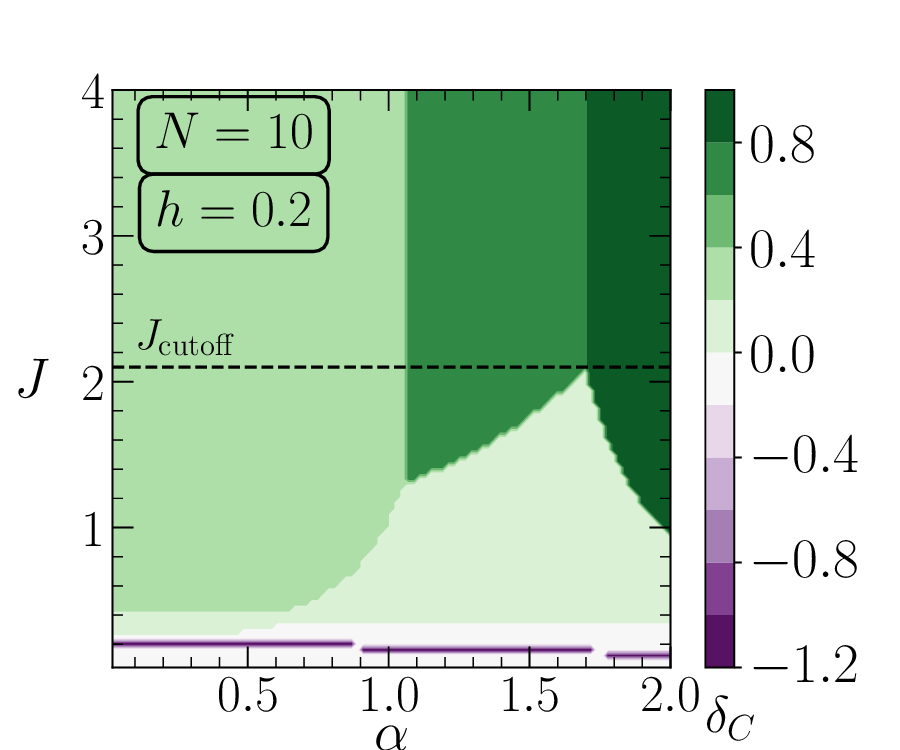}
    \label{fig:ems_10_2}}
  \subfloat[$h=0.4$]{\includegraphics[width=0.25\textwidth]{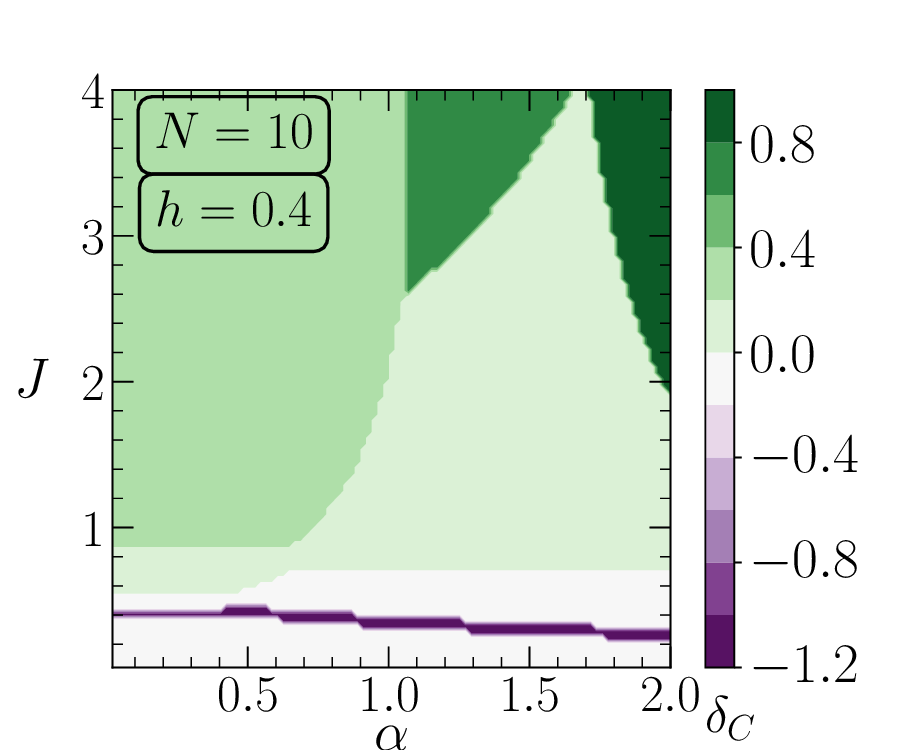}
    \label{fig:ems_10_4}}
  \subfloat[$h=0.9$]{\includegraphics[width=0.25\textwidth]{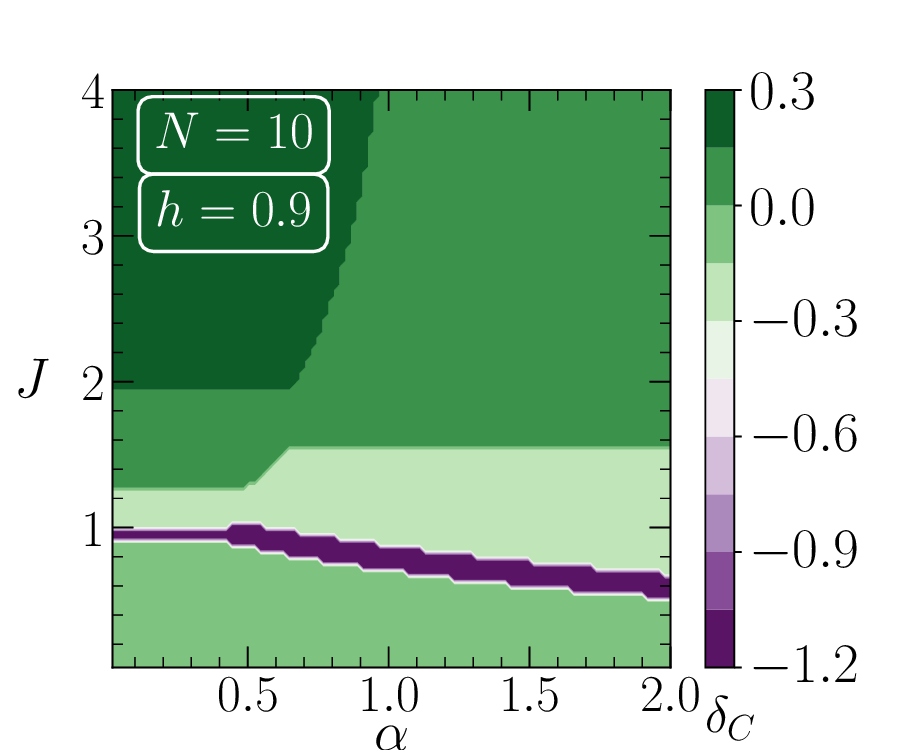}
    \label{fig:ems_10_9}}
\caption{(Color online) Entanglement monogamy score ($\delta_C$) for
various values of two-body ($J$) and three-body ($\alpha$) interactions
of the multiparty quantum spin system given in Eq.~(\ref{eq:ham})
for a
finite spin chain length of $N=10$. Here, plot
(a) is for the absence of an external magnetic field ($h=0$), and plots
(b), (c), and (d) are for the presence of external magnetic field
strengths $h=0.2, 0.4$ and $0.9$ respectively. Monogamous states are represented in
shades of green, and non-monogamous states are represented in shades of
purple. In the absence of an external magnetic field ($h=0$), all the
states are monogamous ($\delta_C > 0$). As $h$ increases, the
percentages of non-monogamous states are 4.19\%, 8.72\%, and 19.01\% for
$h = 0.2, 0.4,$ and $0.9$ respectively.}
\end{figure*}

\begin{figure*}[t]
\label{fig:ems}
  \subfloat[$N=10$, $h=0$, $\forall J$]{\includegraphics[width=0.25\textwidth]{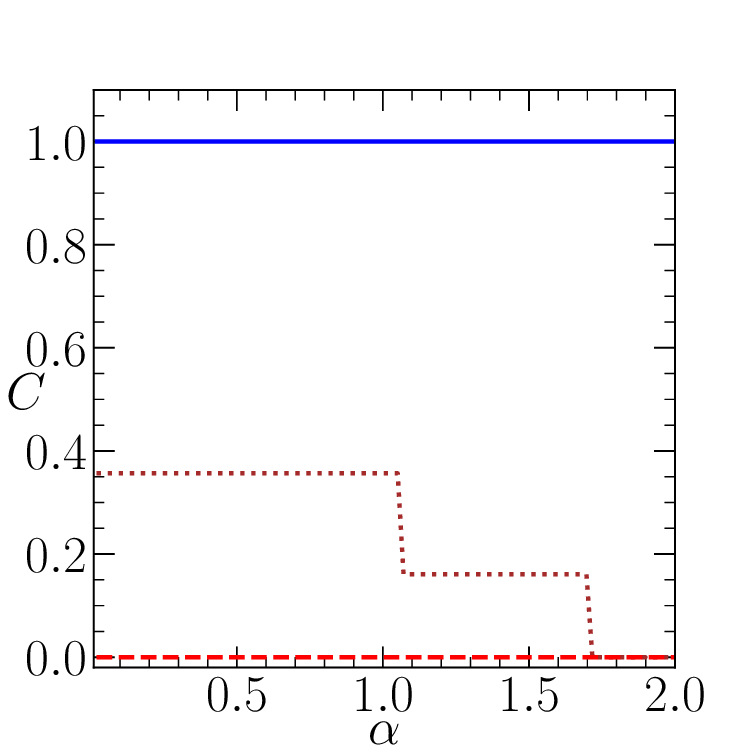}
    \label{fig:concurrence_10_0}}
  \subfloat[$N=10$, $h=0.2$, $J=0.2$]{\includegraphics[width=0.25\textwidth]{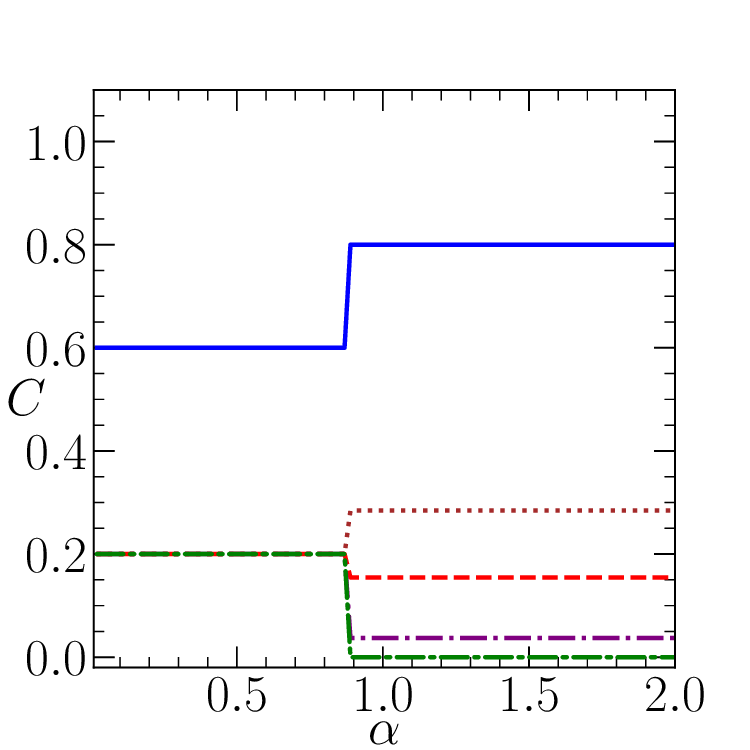}
    \label{fig:concurrence_10_2_J0}}
  \subfloat[$N=10$, $h=0.2$, $J=0.9$]{\includegraphics[width=0.25\textwidth]{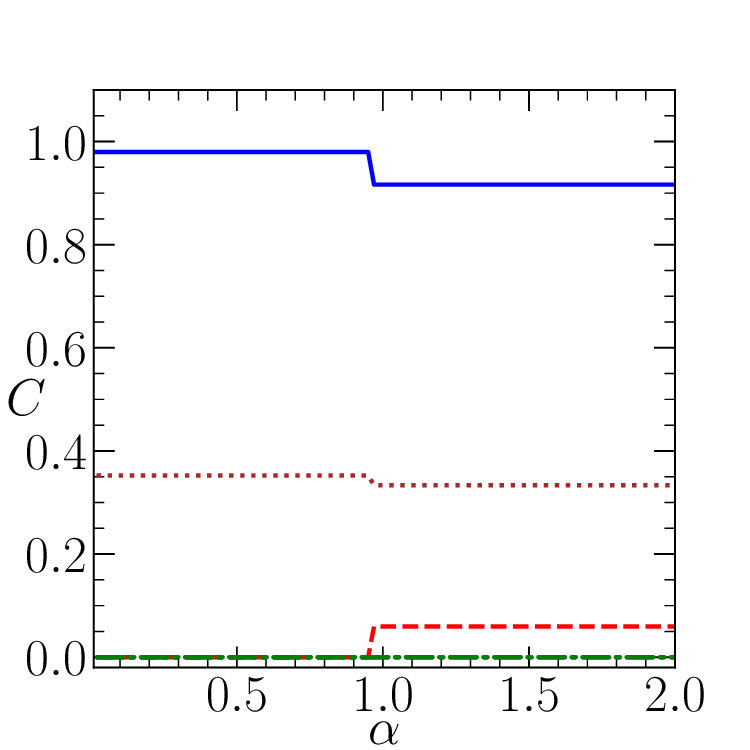}
    \label{fig:concurrence_10_2_J1}}
  \subfloat[$N=10$, $h=0.2$, $J=3$]{\includegraphics[width=0.25\textwidth]{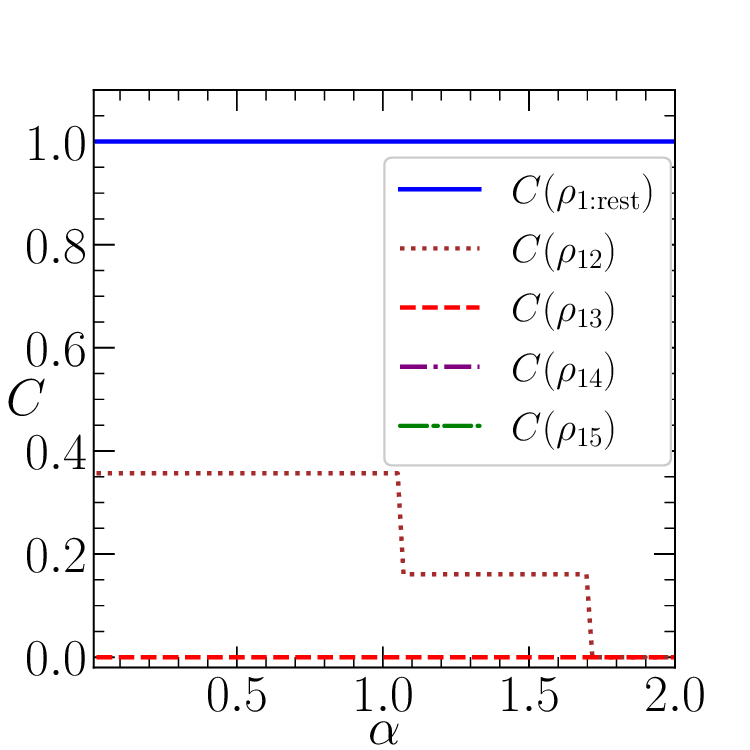}
    \label{fig:concurrence_10_2_J2}}
\caption{(Color online) Whole system concurrence ($C(\rho_{\rm 1 :
rest})$) and subsystem concurrences ($C(\rho_{1k})$; $k=2,3,...,N$) as a
function of three-body interaction $\alpha$ of the multiparty quantum
spin system given in Eq.~(\ref{eq:ham}) for a finite spin chain length
of $N=10$. Here, plot (a) is in the absence of a magnetic field ($h=0$).
Plots (b), (c), and (d) are in the presence of a magnetic field
($h=0.2$) for $J=0.2, 0.9$ and $3$ respectively. In the presence of an
external magnetic field and $J<J_{\rm cutoff}$, as the two-body
interaction ($J$) increases, the subsystem concurrences for the farthest
neighbours of the nodal observer tend to zero, further increase in $J$
will make the next farthest $C$ zero, and the trend continues.  When
$J>J_{\rm cutoff}$, the values of concurrence in the presence of an
external magnetic field are identical to those in the absence of a
magnetic field.}
\end{figure*}

Consider a quantum state with three particles, $\rho_{ABC}$, and a
bipartite quantum correlation measure, $Q$. The state is
said to be monogamous for any $Q$, if
\begin{equation}
    Q(\rho_{A:BC}) \geq Q(\rho_{AB}) + Q(\rho_{AC})
    \label{eq:monogamy_3}
\end{equation}
is satisfied~\cite{coffman2000distributed}. Here, $Q(\rho_{A:BC})$ is
the quantum correlation between subsystem $A$ and subsystems $BC$
together, $Q(\rho_{AB})$ is the quantum correlation between the
subsystems $A$ and $B$ when subsystem $C$ is traced out from
$\rho_{ABC}$, and $Q(\rho_{AC})$ is the quantum correlation between the
subsystems $A$ and $C$ when subsystem $B$ is traced out from
$\rho_{ABC}$. Here, party $A$ is identified as a nodal observer.

Using the above relation, one can define a quantifying measure of
shareability called monogamy score~\cite{bera2012characterization}, for
any quantum correlation measure $Q$, as
\begin{equation} 
\delta_Q = Q({\rho_{A:BC}}) - Q(\rho_{AB}) - Q(\rho_{AC}).
\label{eq:monoscore_3}
\end{equation}
For a given quantum
correlation measure, $Q$, if $\delta_Q \geq 0$
the state $\rho_{ABC}$ is said to be monogamous and if $\delta_Q<0$ the
same is characterized as non-monogamous. When the state is
monogamous, then there exists a restriction on the shareability of $Q$
among the subsystems of $\rho_{ABC}$. The definition of monogamy score
(Eq.~(\ref{eq:monoscore_3})) can be easily extended to an arbitrary $N$-party
quantum state ($\rho_{12 \cdots N}$) as
\begin{equation}
    \delta_Q^j (\rho_{12 \cdots N}) = Q(\rho_{j:{\rm rest}}) - \sum_{k \neq
    j} Q(\rho_{jk}),
    \label{eq:monogamy_N}
\end{equation}
with $j$ being the nodal observer. Without loss of generality, the first
particle ($j=1$) is considered as a nodal observer for the rest of our
study, and we denote the monogamy score in Eq.~(\ref{eq:monogamy_N}) as
$\delta_Q$, ignoring the superscript $j$. In the above equation, $\rho_{jk}$
represents the two-party reduced density matrices, which can be obtained
from the $N$-party density matrix $\rho_{12 \cdots N}$ by tracing out
all the other parties except $j$ and $k$ ($j=1$ and $k=2,3,...,N$). The
schematic representation of various bipartitions across which the
quantum correlation measures are calculated and used in the
equation is given in Fig.~\ref{fig:monogamy}. The various quantum
correlation measures used in Eq.~(\ref{eq:monogamy_N}) to study the
shareability of quantum correlations in our multiparty quantum system of
interest are defined in Appendix~\ref{sec:appendix}.

\begin{figure*}[htpb]
\label{fig:lms}
\centering
  \subfloat[$h=0$]{\includegraphics[width=0.25\textwidth]{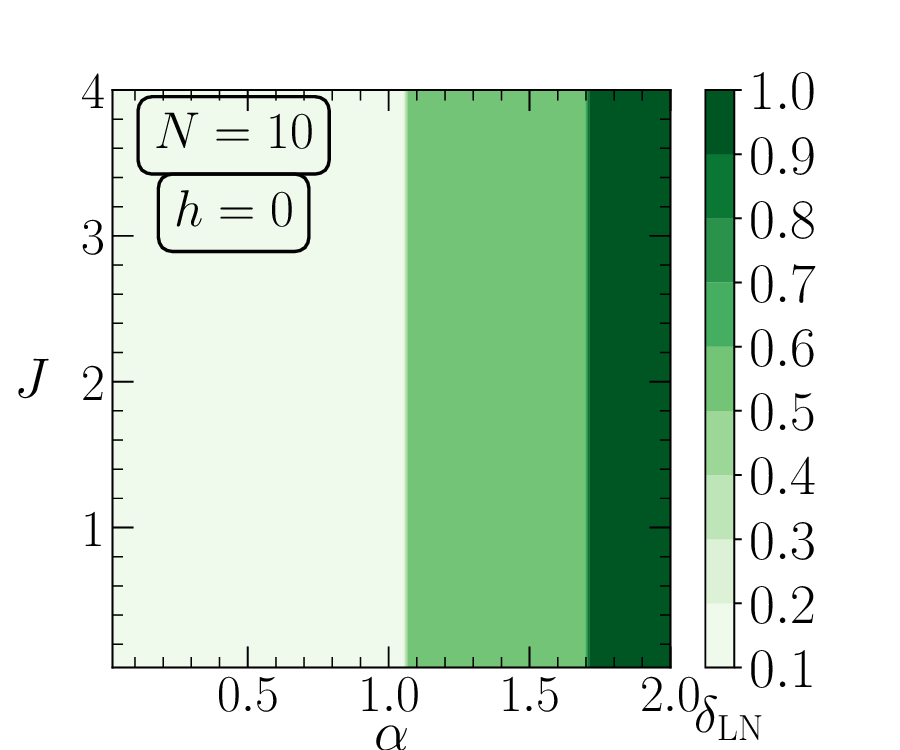}
    \label{fig:lms_10_0}}
  \subfloat[$h=0.2$]{\includegraphics[width=0.25\textwidth]{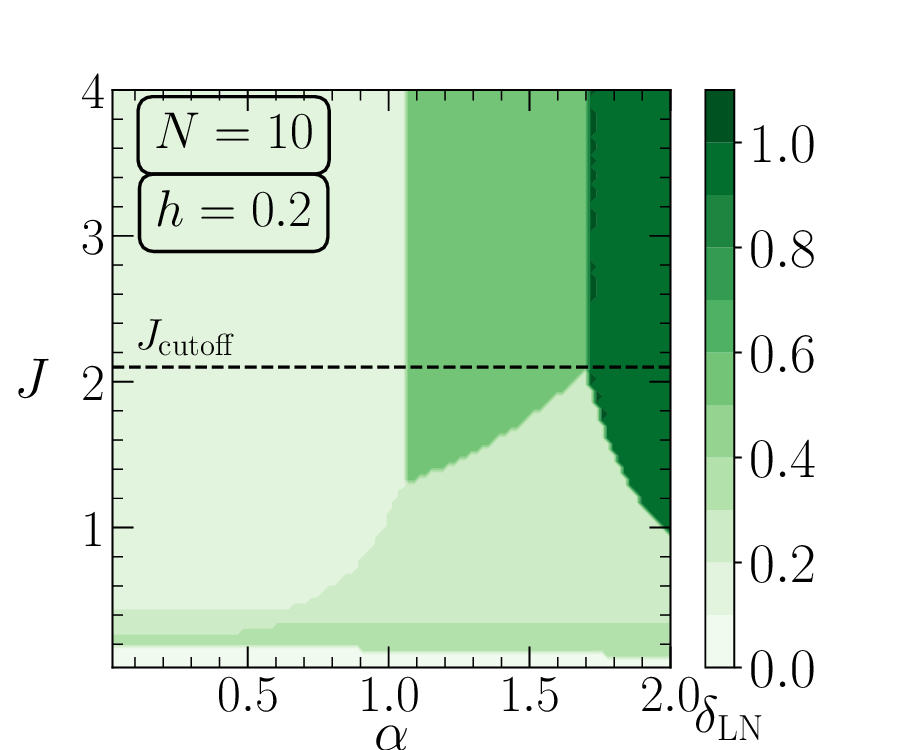}
    \label{fig:lms_10_2}}
  \subfloat[$h=0.4$]{\includegraphics[width=0.25\textwidth]{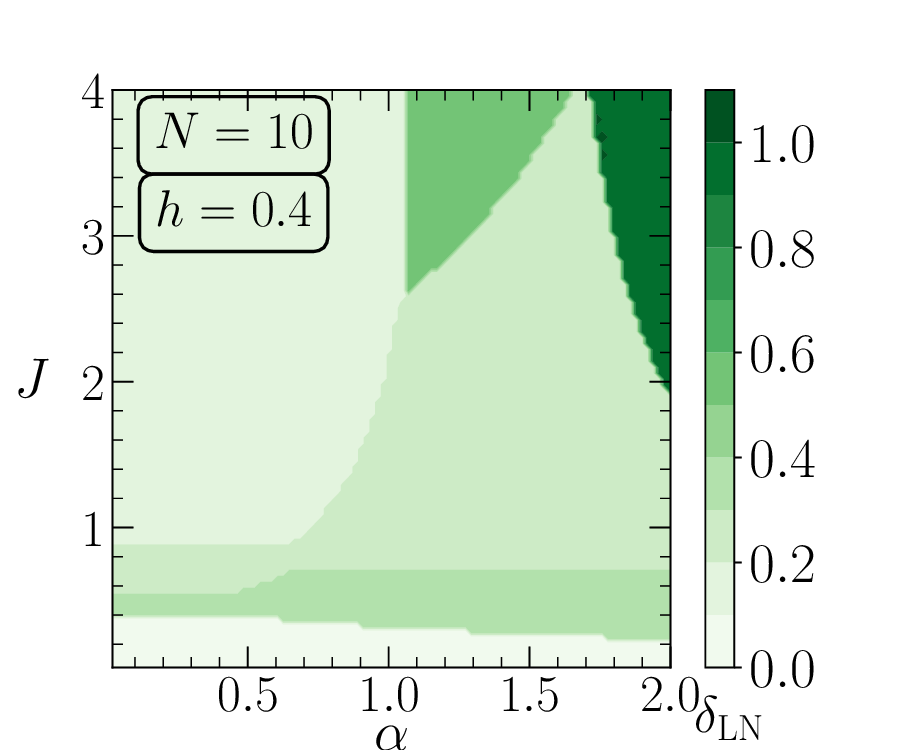}
    \label{fig:lms_10_4}}
  \subfloat[$h=0.9$]{\includegraphics[width=0.25\textwidth]{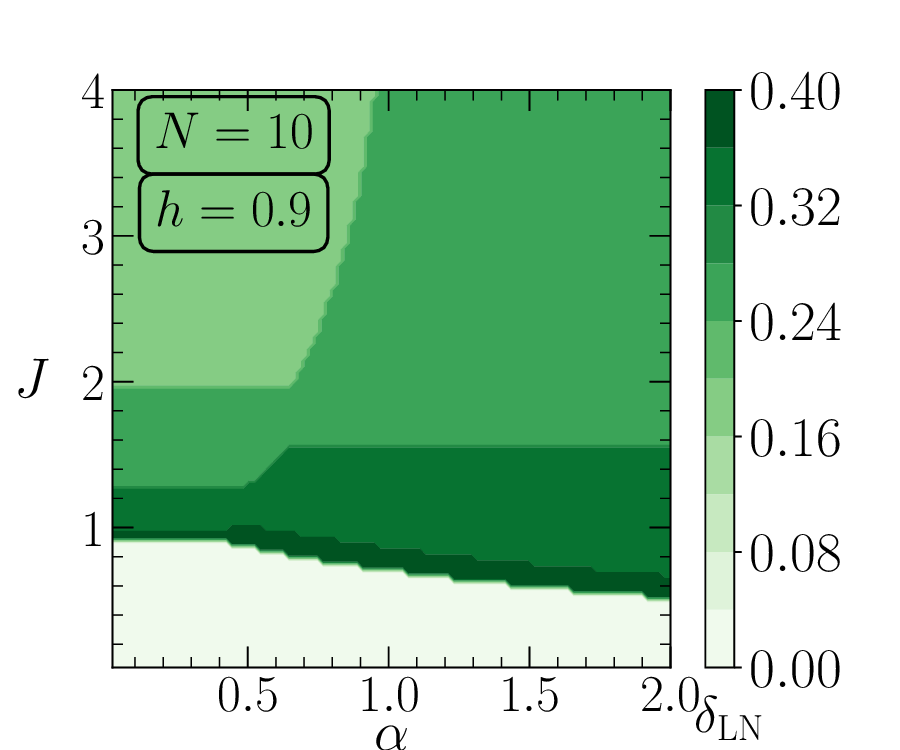}
    \label{fig:lms_10_9}}
\caption{(Color online) Logarithmic negativity monogamy score
($\delta_{\rm LN}$) for various values of two-body ($J$) and three-body
($\alpha$) interactions of the multiparty quantum spin system given in
Eq.~(\ref{eq:ham}) {for a finite spin chain length of $N=10$}. Here, plot
(a) is for the absence of external magnetic field ($h=0$) and plots (b),
(c), and (d) are for the external magnetic field strengths $h=0.2, 0.4$
and $0.9$ respectively.  $\delta_{\rm LN}$ remains monogamous for the
entire range of system parameters $J$, $\alpha$ and $h$.}
\end{figure*}

\section{Shareability of Quantum Correlations in Multiparty Quantum Spin
System}\label{sec:shareability}

To study the shareability of quantum correlations among the subsystems
of a multiparty quantum spin system given in Eq.~(\ref{eq:ham}), the
exact ground state of this Hamiltonian is numerically computed
 for a
finite system length of $N=10$ by varying its system parameters, viz., two-body
interaction ($J$), three-body interaction ($\alpha$), and the external
magnetic field ($h$). 
The restriction on the system length is due to our computational
constraints in numerically handling such many-body quantum systems. We
have chosen to report here the results of $N=10$, however for lower $N$
values, overall conclusions drawn here remain the same.
The corresponding monogamy scores of quantum
correlation measures, which are drawn from both {\em
entanglement-separability} and {\em information-theoretic} kinds have
been evaluated by varying these system parameters.  Here we choose
concurrence~\cite{hill1997entanglement, wootters1998entanglement} and logarithmic
negativity~\cite{plenio2005logarithmic} from the
entanglement-separability, and quantum
discord~\cite{henderson2001classical, ollivier2001quantum} and quantum work
deficit~\cite{oppenheim2002thermodynamical,
horodecki2003local, horodecki2005local, devetak2005distillation} from the
information-theoretic kind of quantum correlation measures. We calculate
the monogamy score of these quantum correlation measures for the system
in Eq.~(\ref{eq:ham}) by varying the system parameters, $J$ and
$\alpha$, in the absence of an external magnetic field ($h=0$), and for
the external magnetic field strengths of $h=0.2$, $0.4$, and $0.9$. The
behaviour of monogamy scores of quantum correlation measures considered
here with respect to these system parameters and external magnetic field
of Hamiltonian in Eq.~(\ref{eq:ham}) is exhibited in this section. We
also consolidate the behaviour of the various bipartite quantum
correlation measures that appear in the definition of monogamy score
with respect to system parameters. 

\subsection{Entanglement-separability Monogamy Scores}

Concurrence, $C$, (for definition see Appendix~\ref{App:concurrence}) is
a well-known measure to quantify the quantum correlations in any
arbitrary two-party quantum systems, like $\rho_{1k}$. By choosing
concurrence as the bipartite quantum correlation measure, the $N$-party
monogamy score (entanglement monogamy score, $\delta_C$) is defined as
\begin{equation}
  \delta_C (\rho_{12 \cdots N}) = C(\rho_{1:{\rm rest}}) - \sum_{k \neq 1}
  C(\rho_{1k}),
    \label{eq:ems}
\end{equation}
with all notions here are carried from Sec.~\ref{sec:monogamy}. The
ground state corresponding to the multiparty quantum spin system with
two- and three-body interactions, given in Eq.~(\ref{eq:ham}),  for $N=10$ will be a
pure state. Hence, concurrence in the 1:rest bipartition is obtained by
$2\sqrt{\det \rho_1}$, where $\rho_1$ is the subsystem of the nodal
observer in the $N$-party quantum state $\rho_{12 \cdots N}$.

For the system given in Eq.~(\ref{eq:ham}), depending on the strength of
external magnetic field, two- and three-body interactions may contribute
to the variation in entanglement monogamy score. However, in the absence
of external magnetic field, $h=0$, (see Fig.~\ref{fig:ems_10_0}), it can
be observed that $\delta_C$ has no dependence on two-body interaction
($J$). Therefore, in the absence of an external magnetic field,
three-body interaction ($\alpha$) is a sole contributor to the
characterization of $\delta_C$. As $\alpha$ increases, $\delta_C$
increases in the form of steps and assumes the value 0.29 for
$\alpha<1$, 0.68 for $1<\alpha<1.7$, and 1 for $\alpha>1.7$. Thus,
beyond $\alpha = 1.7$, for all values of $J$, there exists a maximum
restriction on the shareability of entanglement among the subsystems of
the multiparty quantum spin system with two- and three-body
interactions.  The appearance of multiple sharp transitions at different
values of $\alpha$ in our characterization of monogamy score of quantum
correlations may be due to (i) the finite spin chain and (ii) presence
of quantum phase transitions in the system at those values. However,
note that there exists a quantum phase transition at $\alpha=1$ as
characterized for this system with infinite spins~\cite{liu2012chiral}.
Therefore, as the system size $N$ increases, one of the sharp
transitions may disappear.

When an external magnetic field is applied, i.e., $h\neq0$ in
Eq.~(\ref{eq:ham}), the characterization of $\delta_C$ depends on both
two- and three-body interactions. Further, the influence of two-body
interaction on $\delta_C$ is observed up to a value of $J$, called
$J_{\rm cutoff}$.  For $J>J_{\rm cutoff}$, $\delta_C$ is unaffected by
the introduction of an external magnetic field and the variations in
$\delta_C$ depends only on three-body interaction. For example, in the
case of $h=0.2$ (see Fig.~\ref{fig:ems_10_2}), $\delta_C$ behaves
identical to $h=0$ case (including values) for $J > 2$, hence, $J_{\rm
cutoff} \approx 2$. For $J < J_{\rm cutoff}$, the variation of
$\delta_C$ with respect to $J$ and $\alpha$ is not straightforward. In
this region, for certain values of $J$ and $\alpha$, we observe that
some of the states are non-monogamous, i.e., for those states, $\delta_C
< 0$ (see purple regions in Figs.~\ref{fig:ems_10_2} to
\ref{fig:ems_10_9}). Therefore, for this system, there exists a
parameter range of low $J$ and the whole range of $\alpha$, where there
is no restriction on the shareability of entanglement among the
subsystems of the multiparty quantum spin system under consideration.

When the external magnetic field is further increased to $h=0.4$ and
$0.9$, the region of non-monogamous states increases in the area for low
$J$ and the entire range of $\alpha$. Moreover, the appearance of
$\delta_C<0$ region occurs for slightly increased value of $J$ compared
to $h=0.2$ case, as seen in the bottom part of Figs.~\ref{fig:ems_10_4}
to \ref{fig:ems_10_9}. The white region in Figs.~\ref{fig:ems_10_2} to
\ref{fig:ems_10_9} represents monogamous states with $\delta_C=0$. This
monogamous region keeps increasing its area with the increase in
external magnetic field strength. Note that, the monogamous region above
the purple region decreases with an increase in the external magnetic
field.

\noindent {\em Characterization of concurrence with respect to system
parameters.} As the entanglement monogamy score relation
(Eq.~(\ref{eq:ems})) contains the quantification of entanglement in
different bipartitions of the multiparty spin system under
consideration, we now draw a comparison among these bipartite
entanglements. In other words, we now compare the amount of entanglement
in various two-spin subsystems which includes the common nodal observer
$1$ ($C(\rho_{1k})$) with that of the bipartition between this nodal
observer and the rest of the spins ($C(\rho_{\rm 1:rest})$) in the
system. For simplicity, we refer $C(\rho_{\rm 1:rest})$ as the {\em
whole system entanglement}, and any of the $C(\rho_{1k})$ as the {\em
subsystem entanglement}.

In the absence of an external magnetic field ($h=0$), $\delta_C$ does
not depend on two-body interaction ($J$). Hence, we study the variation
of concurrence with respect to three-body interaction ($\alpha$) alone.
From Fig.~\ref{fig:concurrence_10_0} it is observed that the whole
system correlation ($N=10$) is always maximum ($C(\rho_{\rm 1:rest})=1$)
irrespective of $\alpha$ values. Only for the two-spin subsystems with a
nodal observer and its immediate neighbours, the values of subsystem
entanglement ($C(\rho_{12})$ and $C(\rho_{1N})$) are non-zero. When
$\alpha$ increases, these two entanglements reduce and become zero,
resulting in $\delta_C = C(\rho_{1:{\rm rest}}) = 1$. Therefore, we can
conclude that in the absence of an external magnetic field, the whole
system entanglement always has the maximum value irrespective of the
values of system parameters and the subsystem entanglement reduces to
zero as three-body interaction increases.

In the presence of an external magnetic field, the variation of
concurrence depends on both two- and three-body interactions. Therefore,
we discuss the behaviour of concurrence with respect to three-body
interaction in both $J<J_{\rm cutoff}$ and $J>J_{\rm cutoff}$ regions of
entanglement monogamy score. When the external magnetic field $h$ is
$0.2$, since $J_{\rm cutoff} \approx 2$, we consider two values of $J$
($0.2$ and $0.9$) in the $J<2$ region and one value of $J (=3)$ in the
$J>2$ region. From Figs.~\ref{fig:concurrence_10_0} to
\ref{fig:concurrence_10_2_J2}, one can observe that: (i) The whole
system entanglement is always greater than any of the subsystem
entanglement in both regions of $J$.  (ii) In $J<J_{\rm cutoff}$ region,
as $J$ increases, the subsystem entanglement corresponding to the
farthest spins is the first to tend to zero and later the subsystem
entanglement of the second farthest spin will become zero, this trend
will follow till the behaviour of all the remaining bipartite
entanglement become equivalent to the case of $h=0$. In the $J > J_{\rm
cutoff}$ region, both the whole system and subsystem entanglements
behave identically to the case of absence of an external magnetic field.
(iii) All the subsystem entanglements either increase or decrease with
respect to $\alpha$, depending on the $J$ value. These observations are
true even for $h=0.4$ and $0.9$.  

\noindent{\em Logarithmic negativity monogamy score.} Another quantum
correlation measure considered for our study in the entanglement
separability kind is the logarithmic negativity (for definition see
Appendix \ref{App:LN}). The logarithmic negativity monogamy score,
$\delta_{\rm LN}$, of an $N$-party state is given by
\begin{equation}
	\delta_{\text{LN}} (\rho_{12 \cdots N}) =\text{LN}(\rho_{1:{\rm
  rest}}) - \sum_{k \neq 1}\text{LN}(\rho_{1k}),
	\label{eq:lms}
\end{equation}
with all notions here are carried from Sec.~\ref{sec:monogamy}. For the
Hamiltonian given in Eq.~(\ref{eq:ham}), the variation of logarithmic
negativity monogamy score ($\delta_{\text{LN}}$) with respect to the
system parameters ($J$, $\alpha$ and $h$) for the system size of $N=10$ is consolidated in
Figs.~\ref{fig:lms_10_0} to \ref{fig:lms_10_9}. In the absence of the
magnetic field, $\delta_{\text{LN}}$ depends only on three-body
interaction and attains three distinct values: $ 0.12, 0.57$, and $1$.
These transitions occur at $\alpha \approx 1$ and $1.7$, similar to
$\delta_{C}$. When the external magnetic field of any strength is
introduced in Eq.~(\ref{eq:ham}), the corresponding $\delta_{\text{LN}}$
always remain monogamous ($\delta_{\text{LN}} \geq 0$) irrespective of
the range of the system parameters (see Figs.~\ref{fig:lms_10_0} to
\ref{fig:lms_10_9}). Since $\delta_{\rm LN}$ is monogamous for all
strengths of an external magnetic field, we are not considering the
study of whole system and subsystem logarithmic negativity with respect
to system parameters of the multiparty quantum spin system under
consideration.

\begin{figure*}[htpb]
\centering
  \subfloat[$h=0$]{\includegraphics[width=0.25\textwidth]{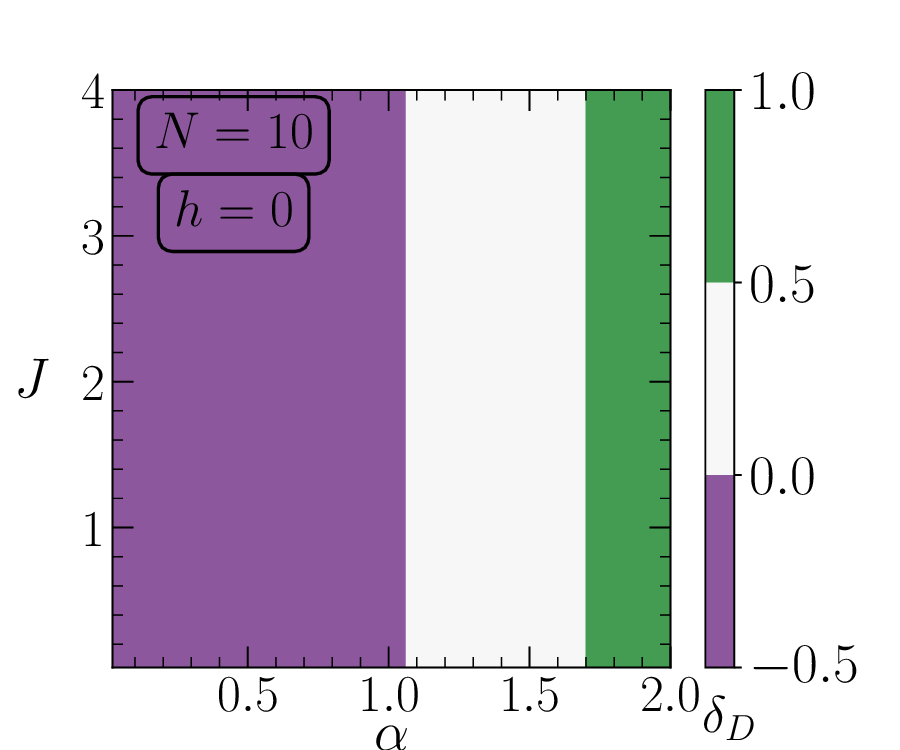}
    \label{fig:dms_10_0}}
  \subfloat[$h=0.2$]{\includegraphics[width=0.25\textwidth]{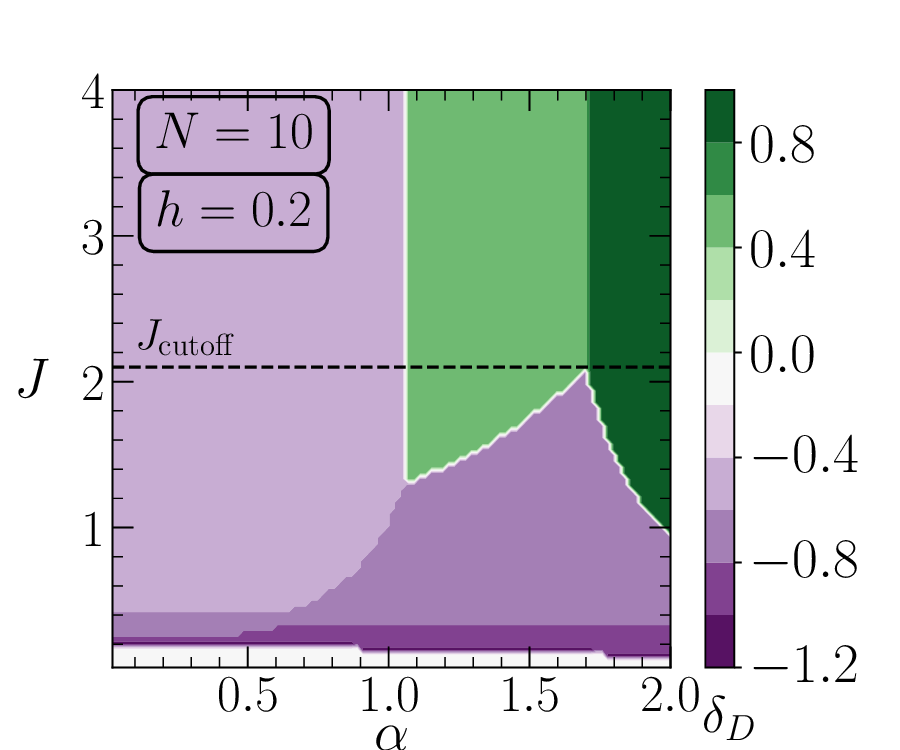}
    \label{fig:dms_10_2}}
  \subfloat[$h=0.4$]{\includegraphics[width=0.25\textwidth]{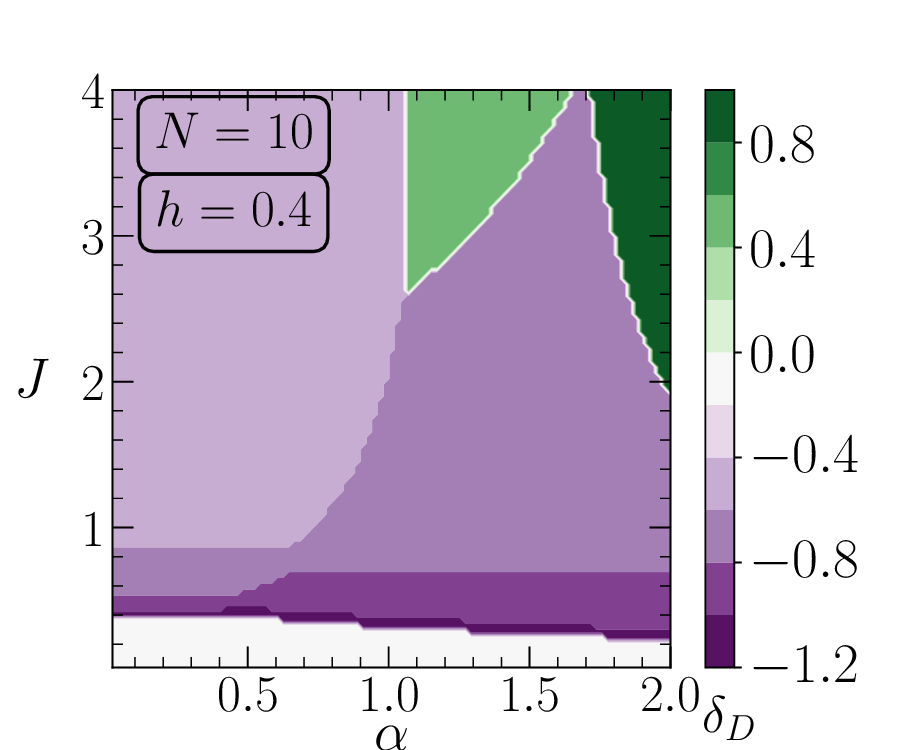}
    \label{fig:dms_10_4}}
  \subfloat[$h=0.9$]{\includegraphics[width=0.25\textwidth]{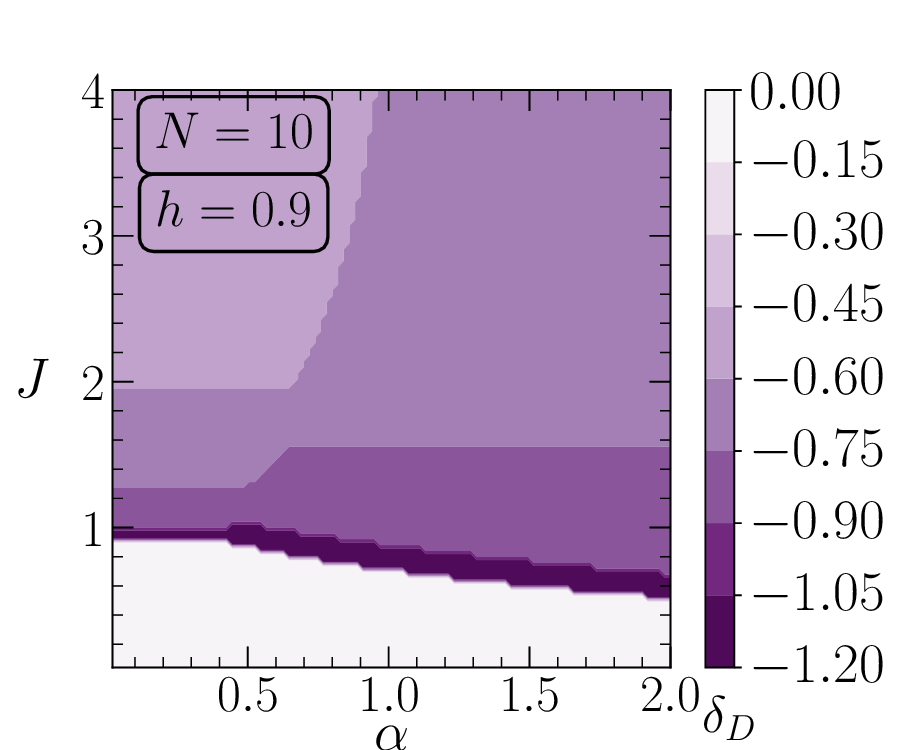}
    \label{fig:dms_10_9}}
\caption{(Color online) Discord monogamy score ($\delta_D$) for various
values of two-body ($J$) and three-body ($\alpha$) interactions of the
multiparty quantum spin system given in Eq.~(\ref{eq:ham}) {for a
finite spin chain length of $N=10$}. Here, plot (a) is for the absence of an
external magnetic field ($h=0$), and plots (b), (c), and (d) are for the
external magnetic field strengths $h=0.2, 0.4$ and $0.9$ respectively. Monogamous
states are represented in shades of green, and non-monogamous states are
represented in shades of purple. In the absence of external magnetic
field ($h=0$), 52.53\% of the states are non-monogamous and as $h$
increases, the percentages of non-monogamous states are 67.27\%, 81.57\%,
and 82.61\% for $h = 0.2, 0.4,$ and $0.9$ respectively.}
\end{figure*}

\begin{figure*}[htpb]
	\label{fig:ems}
	\subfloat[$N=10$, $h=0$, $\forall J$]{\includegraphics[width=0.25\textwidth]{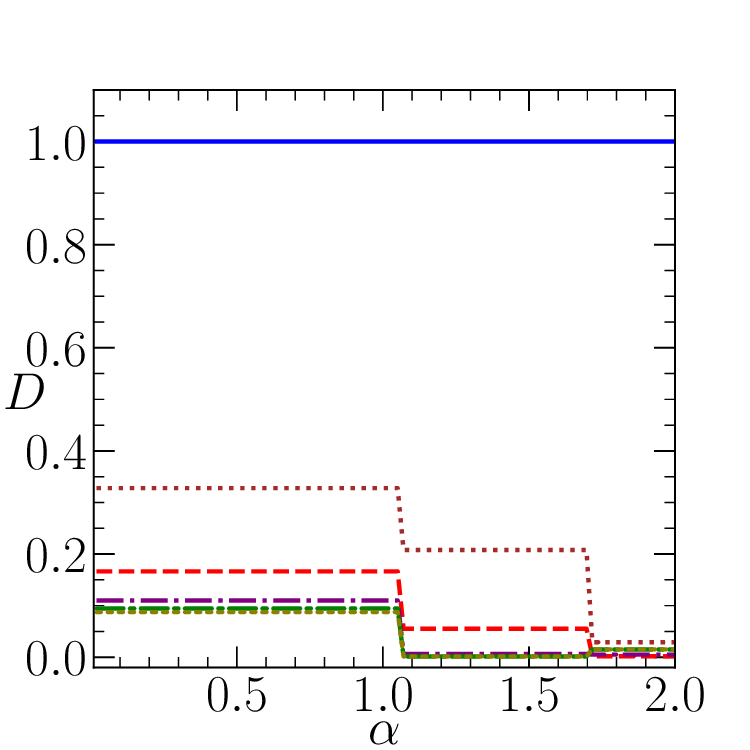}
		\label{fig:discord_10_0}}
	\subfloat[$N=10$, $h=0.2$, $J=0.2$]{\includegraphics[width=0.25\textwidth]{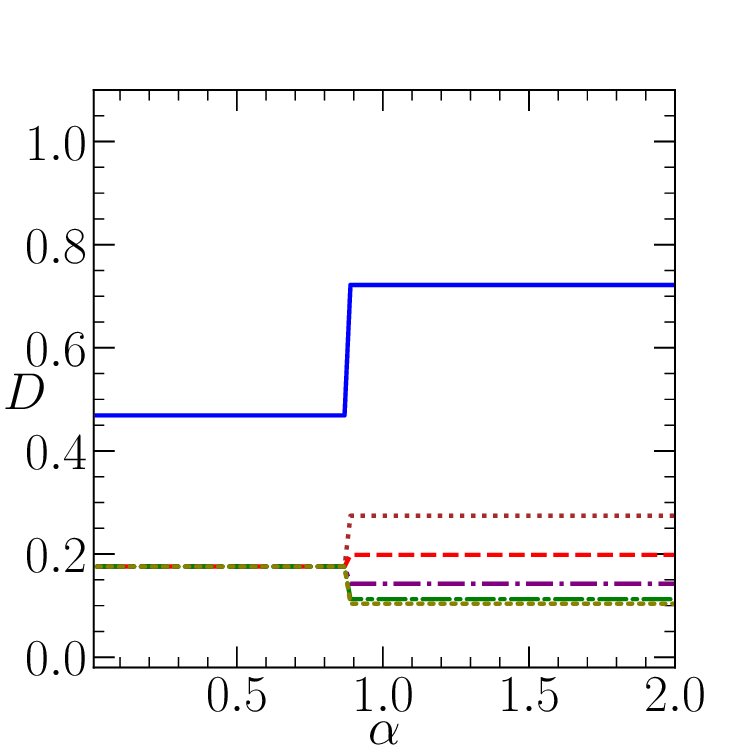}
		\label{fig:discord_10_2_J0}}
	\subfloat[$N=10$, $h=0.2$, $J=0.9$]{\includegraphics[width=0.25\textwidth]{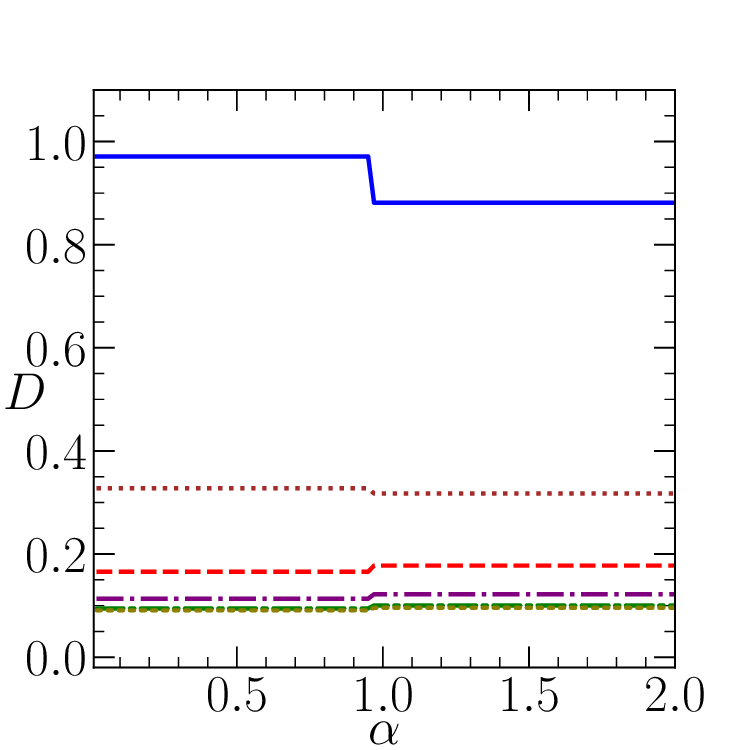}
		\label{fig:discord_10_2_J1}}
	\subfloat[$N=10$, $h=0.2$, $J=3$]{\includegraphics[width=0.25\textwidth]{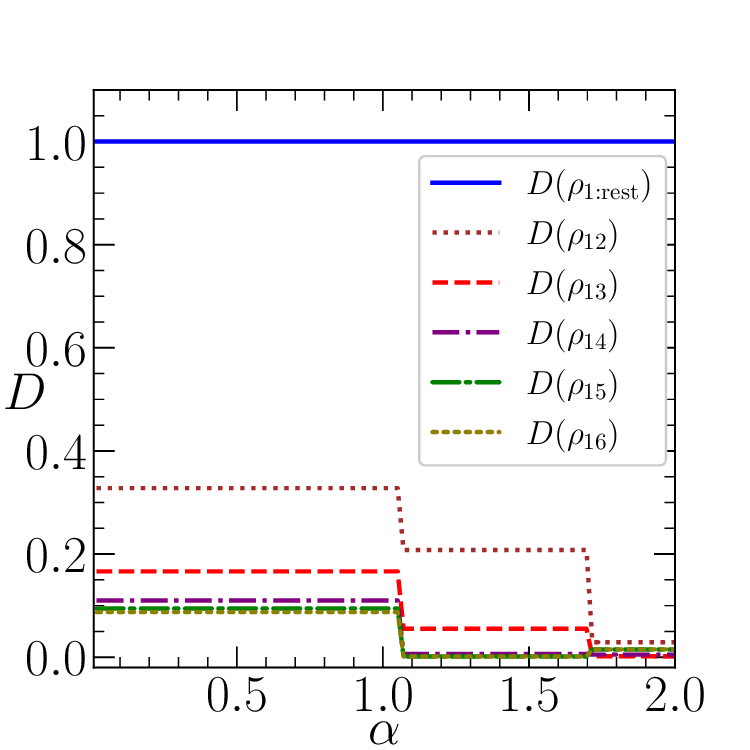}
		\label{fig:discord_10_2_J2}}
  \caption{(Color online) Whole system quantum discord ($D(\rho_{\rm 1 :
  rest})$) and subsystem quantum discords ($D(\rho_{1k})$; $k=2,3,...,N$) as a function of
  three-body interaction $\alpha$ of the multiparty quantum spin system
  given in Eq.~(\ref{eq:ham}) {for a finite spin chain length
  of $N=10$}.  Here, plot (a) is in the absence of an external magnetic
  field ($h=0$). Plots (b), (c), and (d) are in the presence of an
  external magnetic field ($h=0.2$) for $J=0.2, 0.9$ and $3$
  respectively. When
  $J>J_{\rm cutoff}$, the values of quantum discord in the presence of
  the magnetic field are identical to those in the absence of an
  external magnetic field.}
\end{figure*}

\subsection{Information-theoretic Monogamy Scores}

We now study the behaviour of monogamy scores of information-theoretic
quantum correlation measures for the multiparty quantum spin system with
two- and three-body interactions given in Eq.~(\ref{eq:ham}) for a
finite system length of $N=10$.  Let us
first consider quantum discord (see Appendix \ref{App:discord}) as a
information-theoretic quantum correlation measure and the corresponding
$N$-party discord monogamy score ($\delta_D$) is defined as
\begin{equation}
	\delta_D (\rho_{12 \cdots N}) = D(\rho_{1:{\rm rest}}) - \sum_{k \neq 1}
	D(\rho_{1k}),
	\label{eq:dms}
\end{equation}
with all notions here are carried from Sec.~\ref{sec:monogamy}. From
Fig.~\ref{fig:dms_10_0}, it can be observed that when the external
magnetic field is absent ($h=0$), only three-body interaction ($\alpha$)
contribute to the variation of discord monogamy score. The discord
monogamy score  $\delta_D$ has the value $-0.49$ in the region $
0<\alpha<1 $ (purple region), indicating that the multiparty state in
this region have no restriction on the sharing of quantum correlations
in the form of quantum discord among its subsystems.  As $\alpha$
increases, $\delta_D$ assumes the values $0.45$ for $1 < \alpha < 1.7$
and $0.88$ for $\alpha > 1.7$, indicating that there exists a
restriction on the sharing of quantum discord among the subsystems of
Eq.~(\ref{eq:ham}) when $\alpha>1$.

When we introduce the external magnetic field, the discord monogamy
score ($\delta_D$) depends on both two- and three-body interactions,
similar to the entanglement monogamy score. Further, the influence of
two-body interaction on $\delta_D$ is observed only until $J_{\rm
cutoff} \approx 2$. For example, in the case of $h=0.2$ (see
Fig.~\ref{fig:dms_10_2}), $\delta_D$ behaves identically to the case of
absence of an external magnetic field for $J > 2$. For $J < J_{\rm
cutoff}$, the variation of $\delta_D$ with respect to $J$ and $\alpha$
is not straightforward. However, in the $J < J_{\rm cutoff}$ region, we
can observe that most of the states are non-monogamous. As we increase
the magnetic field strength ($h$), the non-monogamous region with
respect to both $J$ and $\alpha$ increases (see Figs.~\ref{fig:dms_10_0}
and \ref{fig:dms_10_2}). The subsequent increase of external magnetic
field strength ($h$) to $0.4$ and $0.9$ leads to further increase in the
region of non-monogamous states ($\delta_D < 0$), as seen in
Figs.~\ref{fig:dms_10_4} and \ref{fig:dms_10_9} respectively. Note that
in Figs.~\ref{fig:dms_10_2} to \ref{fig:dms_10_9}, the region below the
non-monogamous states are monogamous with $\delta_D=0$, and this region
keeps increasing with an increase in the magnetic field strength.
Therefore, in the presence of external magnetic field, for most of the
values of two-body interaction $J$ and the whole range of three-body
interaction $\alpha$, there is no restriction on shareability of quantum
discord among the subsystems of the multiparty quantum spin system under
consideration.

\noindent {\em Characterization of quantum discord with respect to
system parameters.} We now compare the whole system and the subsystem
correlations using quantum discord as the quantum correlation measure.
In the absence of an external magnetic field, we study the behaviour of
quantum discord with respect to three-body interaction ($\alpha$), as
discord monogamy score ($\delta_D$) is independent of two-body
interaction $J$.  From Fig.~\ref{fig:discord_10_0}, it can be observed
that the whole system quantum discord, i.e., $N=10$, ($D(\rho_{1:{\rm rest}})$) always
has the maximum value, and all the subsystem quantum discords
($D(\rho_{1k})$) exist in the entire range of $\alpha$. One has to note
here that, contrary to entanglement, subsystem quantum discords will
never vanish for various strengths of system parameters (see
Figs.~\ref{fig:discord_10_0} to \ref{fig:discord_10_2_J2}), hence
allowing the system to be more non-monogamous when compared to
entanglement monogamy score.

In the presence of an external magnetic field $h=0.2$, since we observed
$J_{\rm cutoff} \approx 2$ similar to that of $\delta_C$, we study
quantum discord for the same values of $J$ ($0.2, 0.9$ and $3$). From
Fig.~\ref{fig:discord_10_2_J0} and \ref{fig:discord_10_2_J1}, in the
$J<J_{\rm cutoff}$ region, we observe that all the subsystem quantum
discords ($D(\rho_{1k})$) exist and can increase or decrease with
respect to $\alpha$ values. In $J > J_{\rm cutoff}$ region, the whole
system and subsystem quantum discords are equivalent to the case of
absence of an external magnetic field (see
Fig.~\ref{fig:discord_10_2_J2}). Such behaviour of quantum discord is
also observed for $h=0.4$ and $0.9$.

\begin{figure*}[htpb]
\centering
  \subfloat[$h=0$]{\includegraphics[width=0.25\textwidth]{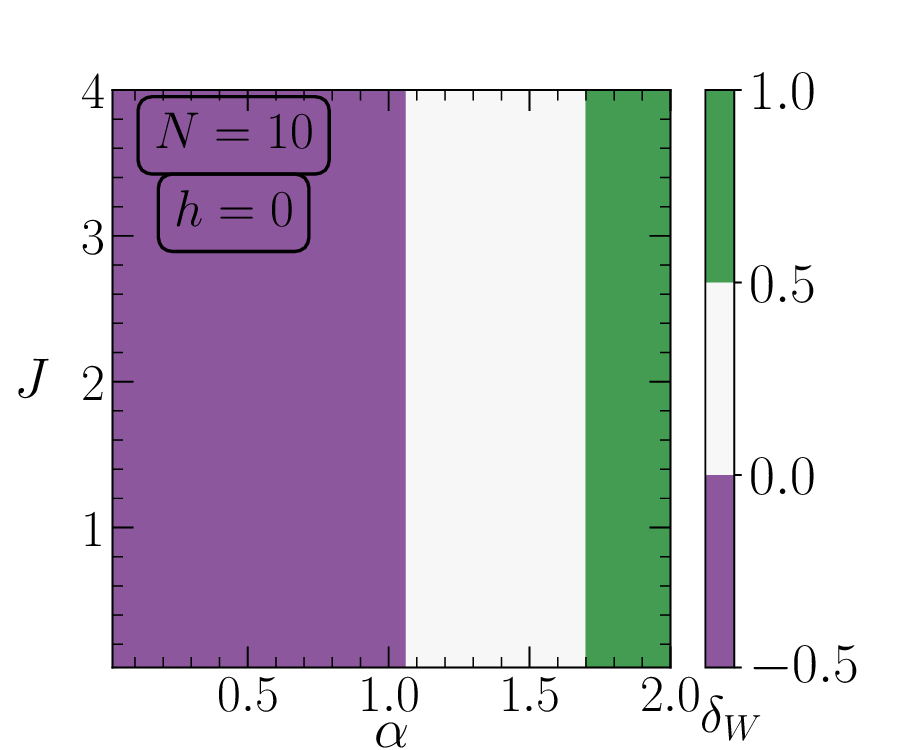}
    \label{fig:wms_10_0}}
  \subfloat[$h=0.2$]{\includegraphics[width=0.25\textwidth]{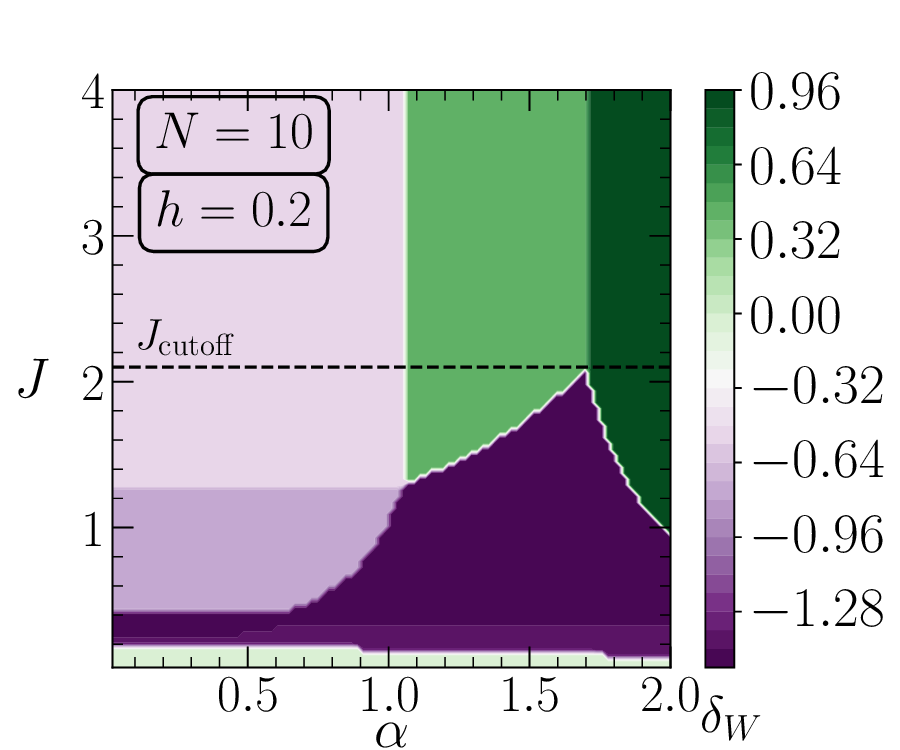}
    \label{fig:wms_10_2}}
  \subfloat[$h=0.4$]{\includegraphics[width=0.25\textwidth]{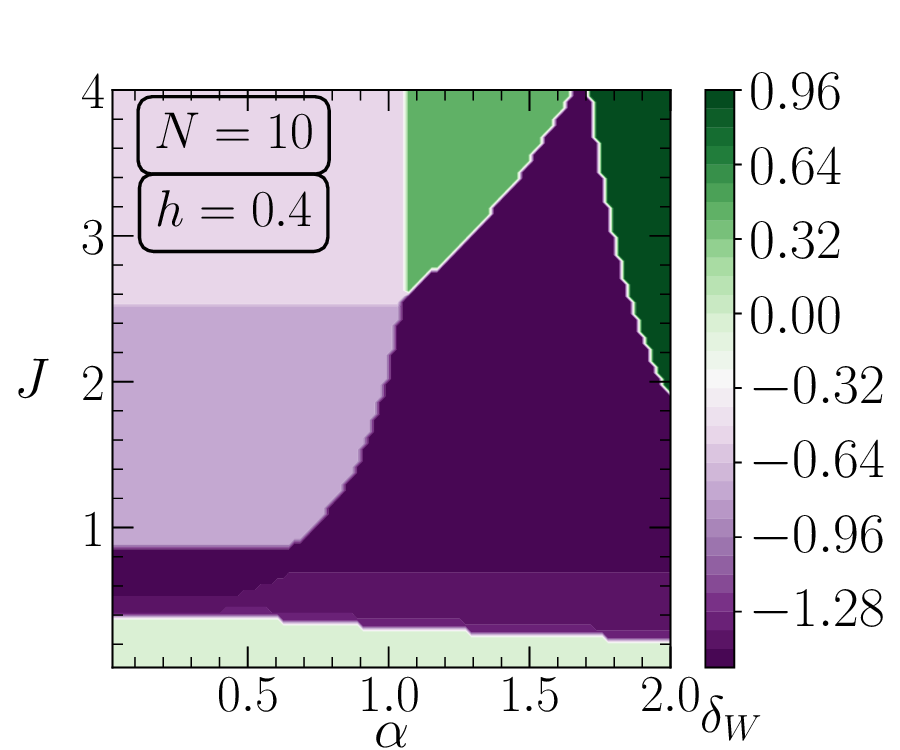}
    \label{fig:wms_10_4}}
  \subfloat[$h=0.9$]{\includegraphics[width=0.25\textwidth]{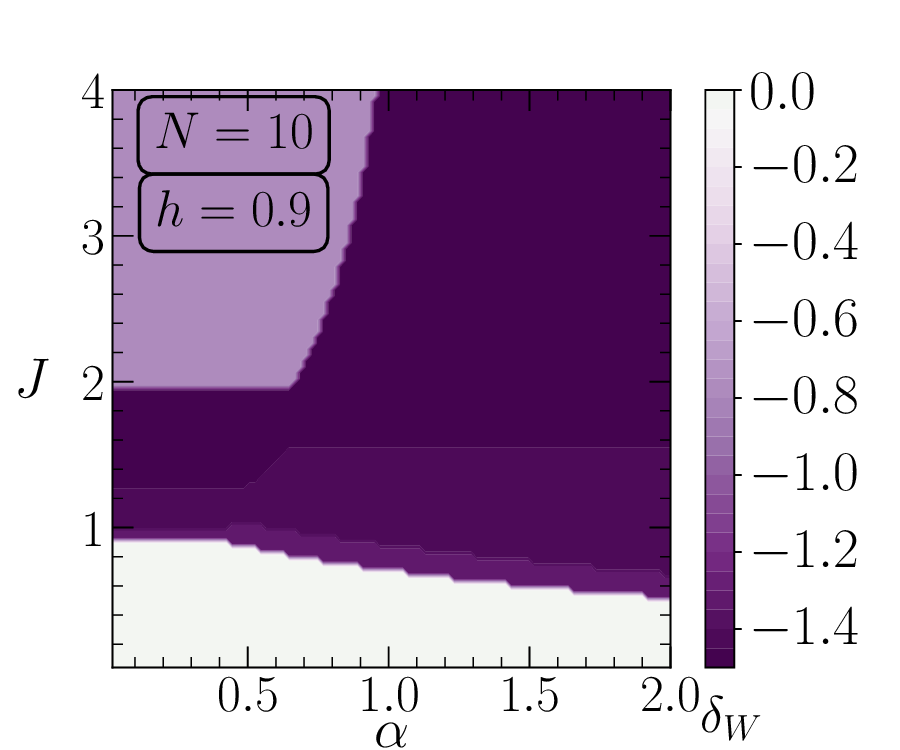}
    \label{fig:wms_10_9}}
\caption{(Color online) Work deficit monogamy score ($\delta_W$) for
various values of two-body ($J$) and three-body ($\alpha$) interactions
of the multiparty quantum spin system given in Eq.~(\ref{eq:ham})
{ for a
finite spin chain length of $N=10$}. Here,
plot (a) is for the absence of external magnetic field ($h=0$), and
plots (b), (c), and (d) are for the external magnetic field strengths
$h=0.2, 0.4$ and $0.9$ respectively. Monogamous states are represented
in shades of green, and non-monogamous states are represented in shades
of purple. In the absence of an external magnetic field ($h=0$), 52.53\%
of the states are non-monogamous and as $h$ increases, the percentages
of non-monogamous states are 67.27\%, 81.57\%, and 82.61\% for $h = 0.2,
0.4,$ and $0.9$ respectively.}
\end{figure*}

\noindent{\em Work deficit monogamy score.} We now consider quantum work
deficit (see Appendix \ref{App:work}) as another example for
information-theoretic quantum correlation measure. The corresponding
$N$-party work deficit monogamy score ($\delta_W$) is given by
\begin{equation} 
\delta_{W} (\rho_{12 \cdots N}) =W(\rho_{1:{\rm rest}})
- \sum_{k \neq 1}W(\rho_{1k}),  
\label{eq:wms} 
\end{equation}
with all notions here are carried from Sec.~\ref{sec:monogamy}. The work
deficit monogamy score ($\delta_{W}$) for the system under consideration
(see Eq.~(\ref{eq:ham})),  with $N=10$, behaves similar to discord monogamy score
($\delta_{D}$) for the entire range of system parameters.  From
Fig.~\ref{fig:wms_10_0} it can be observed that, in the absence of
magnetic field ($h=0$), only three-body interaction ($\alpha$)
contribute to the variation of $\delta_W$, and in the presence of $h$,
$\delta_W$ depends on both $J$ and $\alpha$, which is similar to
$\delta_D$. The existence of $J_{\rm cutoff}$ for $h=0.2$ case (see
Fig.~\ref{fig:wms_10_2}) and the variation of $\delta_W$ in the $J <
J_{\rm cutoff}$ and $J > J_{\rm cutoff}$ regions are also same as
discord monogamy score. Upon further investigating the whole system
correlation and subsystem correlations, it is observed that they also
show similar variations as that of the quantum discord. Therefore, we
can conclude that both of these information-theoretic measures exhibit
same characteristics for shareability of quantum correlations among the
subsystems of the multiparty quantum spin system given in
Eq.~(\ref{eq:ham}) for any choice of strengths of system parameters and
external magnetic field.

\section{Percentage of non-monogamous states}\label{sec:percentage}

In the previous section, we observed that the multiparty spin system
under consideration exhibited both monogamous and non-monogamous
signatures for different choices of quantum correlation measures for
various ranges of system parameters ($J$, $\alpha$, and $h$)
{ for a
finite spin chain length of $N=10$}.
Therefore, now we compare the percentage of non-monogamous states for
all the quantum correlation measures considered here. For such a study,
we generate $10^4$ states in the allowed ranges of $J$ and $\alpha$ for
a fixed value of an external magnetic field.

{\em Entanglement-separability monogamy scores}: The entanglement
monogamy score ($\delta_C$) exhibits a monogamous nature for all the
states in the absence of an external magnetic field, and as the magnetic
field increases, some of the states become non-monogamous. The
percentage of states that are non-monogamous with respect to $\delta_C$
in the presence of external magnetic field strengths of $0.2$, $0.4$,
and $0.9$ are
4.19\%, 8.72\% and 19.01\% respectively (see Table~\ref{tab:power}).
  However, logarithmic negativity monogamy score ($\delta_{\text{LN}}$)
  always remains monogamous, irrespective of changes in the values of
  the system parameters and the external magnetic field.

{\em Information theoretic monogamy scores}: When discord monogamy score
($\delta_D$) is considered, we observe that, in the absence of an
external magnetic field, 52.5\% of the states exhibit non-monogamous
behaviour. As the magnetic field is increased to $h=0.2$, $0.4$ and
$0.9$, the percentage of non-monogamous states increased to 67.27\%,
81.57\% and 82.61\% respectively (see Table~\ref{tab:power}). In the
   case of work deficit monogamy score $\delta_W$, the percentages of
   non-monogamous states are observed to be equivalent to that of
   discord monogamy score. However, it must be noted that even though
   the percentage of non-monogamous states in the cases of $\delta_D$
   and $\delta_W$ are equal, the corresponding values of $\delta_D$ and
   $\delta_W$ for a given set of system parameters ($J$, $\alpha$, and
   $h$) are not equal. 

Recently, it is shown that a quantum correlation measure that is
initially non-monogamous for a multiparty quantum state can be made
monogamous by considering an increasing function of the same measure,
such as integral powers of quantum correlation
measures~\cite{salini2014monotonically}. Such increasing functions of
quantum correlation measures maintain all the properties of the original
measures, such as monotonicity under local operations. Here, we define
the monogamy scores for the integral powers of quantum correlation
measures as
\begin{equation}
	\delta_{Q^m} = Q^m(\rho_{1:{\rm rest}}) - \sum_{k \neq 1} Q^m(\rho_{1k}),
\end{equation}
with all notions here are carried from Sec.~\ref{sec:monogamy}. When the
multiparty state $\rho_{12 \cdots N}$ exhibits non-monogamous nature
($\delta_Q < 0$) for a quantum correlation measure $Q$, there exists a
positive integer $m$ for which $\delta_{Q^m} >
0$~\cite{dhar2017monogamy}.  We now calculate the monogamy score using
the integral powers of quantum correlation measures for the multiparty
quantum spin system given in Eq.~(\ref{eq:ham}) and observe the value of
$m$ for which the non-monogamous states of this system will become
monogamous. We consider the same data of $10^4$ states for which the
percentages of non-monogamous states has been obtained for the allowed
ranges of system parameters and external magnetic field.  In the
entanglement-separability kind of quantum correlation measures, the
logarithmic negativity monogamy score is always monogamous. Therefore,
for the trivial case of $\delta_{\rm LN}$, the integral power $m$ is 1.
For the case of entanglement monogamy score, we have found that all the
states of the given system will become monogamous when $m=2$.  This
result for squared concurrence concurs with the findings of Coffman et
al.~\cite{coffman2000distributed}, viz., the squared concurrence is
always monogamous for any multiparty quantum state. In the
information-theoretic kind of quantum correlation measures, both discord
monogamy score and work deficit monogamy scores will become monogamous
for the integral power of $m=3$. All these results are summarized in the
Table~\ref{tab:power}.

\begin{table}[ht]
    \caption{\label{tab:power} Percentage of states showing
    non-monogamous nature ($\delta_Q <0$) for different quantum
    correlation measures ($Q$) with varying external magnetic field
    ($h$)
    across the multiparty quantum spin system with two- and three-body
    interactions  given in Eq.~(\ref{eq:ham}) for system size of $N=10$. It is clear that at higher magnetic field strengths,
    the quantum correlation measures have more restriction on
    shareability when compared to low magnetic field strengths. The
    integral power $m$ of quantum correlation measure for which all
    states become monogamous ($\delta_{Q^m} >0$) is given in the last column.}
    \small 
    \setlength{\tabcolsep}{4pt} 
    \begin{tabularx}{\columnwidth}{*{6}{>{\centering\arraybackslash}X}}
    \hline
        \multirow{2}{*}{$\delta_Q$} & \multicolumn{4}{c}{\% of non-monogamous
        states} & \multirow{2}{*}{$m$} \\[0.2cm]
        & $h$=0 & $h$=0.2 & $h$=0.4 & $h$=0.9 & \\
        \hline
        $\delta_C$ & 0 & 4.19 & 8.72 & 19.01 & 2 \\
        $\delta_{\text{LN}}$ & 0 & 0 & 0 & 0 & 1 \\
        $\delta_D$ & 52.53 & 67.27 & 81.57 & 82.65 & 3 \\
        $\delta_W$ & 52.53 & 67.27 & 81.57 & 82.65 & 3 \\ \hline
    \end{tabularx}
\end{table}

\section{Finite Size Scaling}\label{sec:finite}

{
We observe two sharp transitions in the values of all the monogamy
scores of quantum correlation measures, considered in the previous section, with respect to three-body
interaction $\alpha$, in the absence of external applied magnetic field
$h$ across the system given in Eq.~(\ref{eq:ham}). In the presence of an external magnetic field of strength
$h=0.2$, there exists a value of two-body interaction strength, $J_{\rm cutoff}$, above which these sharp transitions sustain, identical to the
$h=0$ case. For two-body interaction strengths $J < J_{\rm cutoff}$, the sharp
transitions show varied behaviour with respect to $J$. For higher $h$ the transitions of monogamy scores with respect to $\alpha$ becomes more involved.

The quantum phase transition of this multiparty quantum spin system with two- and three-body interactions at with infinite chain length for which Hamiltonian given in Eq.~(\ref{eq:ham}) is investigated in~\cite{lou2004quantum}. The first sharp transition $\alpha_{c_1}$ of the monogamy scores
occurs around $\alpha_c \approx 1$ which is an indicator of the quantum phase
transition present in this system. Our numerical analysis of monogamy scores finite
spin chain lengths of $N=8, 10$ and $12$, the first sharp transition
occurs at $\alpha_{c_1}=1.09, 1.06$ and $1.04$ respectively. We can identify
that as the system size $N$ increases, the scaling for $\alpha_{c_1}$ given
as
\begin{equation}
  \alpha_{c_1} \sim \alpha_c + 0.4556\,\, e^{-0.2027 N}.
\end{equation}
Clearly, as $N\to \infty$, $\alpha_{c_1}$ will tend to $\alpha_c=1$. The second sharp
transitions of monogamy scores with respect to $\alpha$ occur at
$\alpha_{c_2} = 2.62, 1.71$ and $1.42$ for $N=8, 10$ and $12$
respectively. The second sharp transition scales with system size $N$ as
\begin{equation}
  \alpha_{c_2} \sim \alpha_c +  31.4705\,\, e^{-0.3718 N}.
\end{equation}
As the system size $N$ increases, this value of $\alpha_{c_2}$ will
also converge to the quantum phase transition point of $\alpha_c=1$. These
transition points remain the same irrespective of quantum correlation measures considered in the monogamy score. 
}

\section{Conclusion}\label{sec:conclusion}

The quantum correlation measures, both from the
entanglement-separability and the information-theoretic kinds, can
exhibit monogamous or non-monogamous nature depending on the multiparty
quantum system. Monogamy score, defined using the monogamy relation, is
a quantitative indicator that characterizes the extent to which quantum
correlations can be shared among the subsystems of a multiparty quantum
system. Here, we use the monogamy score of several quantum correlation
measures to quantify their shareability among the subsystems of a
multiparty quantum spin system, { with 10 spins in the chain}, which contains both two- and three-body
interactions, $J$ and $\alpha$ respectively. We also have considered the
cases of with and without the influence of external magnetic field on
this system. 

In the entanglement-separability kind of quantum correlation measures,
we considered concurrence and logarithmic negativity for the calculation
of monogamy scores. In the absence of an external magnetic field, the
entanglement monogamy score ($\delta_C$) depends only on three-body
interaction.  When an external magnetic field is introduced, $\delta_C$
depends on both two- and three-body interactions. For an external
magnetic field strength of
0.2, there exists a value of $J$, called $J_{\rm cutoff}$, above which
  the value of $\delta_C$ is unaffected by the application of external
  magnetic field. All states are monogamous in the absence of an
  external magnetic field, and some of the states become non-monogamous
  as the magnetic field is introduced. It is observed that for the
  external magnetic field strength $h=0.2$, 4.19\% of the states are
  non-monogamous, and this percentage increases to 19.01\% with the
  increase in $h$ to 0.9. When the monogamy score is calculated for
  integral powers of the quantum correlation measure, all states became
  monogamous for concurrence raised to the integral power of 2.  In case
  of logarithmic negativity monogamy score ($\delta_{\rm LN}$), it
  depends only on three-body interaction when an external magnetic field
  is absent, whereas in the presence of an external magnetic field,
  $\delta_{\rm LN}$ depends on both two- and three-body interaction.
  Similar to the case of $\delta_C$, a $J_{\rm cutoff}$ exists for the
  magnetic field strength $h=0.2$ above which $\delta_{\rm LN}$ is
  unaffected by the magnetic field. However, unlike the case of
  $\delta_C$, logarithmic negativity remains monogamous for all ranges
  of system parameters and external magnetic field strengths.

In the information-theoretic type of quantum correlation measures, we
considered quantum discord and quantum work deficit for the calculation
of monogamy scores. Discord monogamy score ($\delta_D$) depends only on
three-body interaction in the absence of an external magnetic field.
When an external magnetic field is introduced, $\delta_D$ depends on
both two- and three-body interactions. For an external magnetic field
strength of $h=0.2$, $J_{\rm cutoff} \approx 2$ exists for $\delta_D$
similar to the case of $\delta_C$. However, 52.53\% of the states showed
non-monogamous signature with respect to quantum discord, even in the
absence of an external magnetic field. In the presence of an external
magnetic field $h=0.2$, 67.27\% of the states showed non-monogamous
behaviour with respect to quantum discord, and with the increase in the
magnetic field strength $h$ to 0.9, 82.65\% of the states became
non-monogamous in the entire range of system parameters. When the
monogamy score was calculated using integral powers of the quantum
correlation measure, all states became monogamous for quantum discord
raised to the power 3.  The characteristic behaviour of bipartite
quantum work deficit, monogamy scores of quantum work deficit
($\delta_W$) and integral power of quantum work deficit ($\delta_{W^m}$)
are exactly same as the corresponding quantities of quantum discord.
However, their values were found to be not equal.

{ The following generic conclusions could be drawn for the system in Eq.~(\ref{eq:ham}): (i) The system exhibits maximum monogamy score for weaker strengths of external applied magnetic field and higher strengths of two- and three-body interactions irrespective of the choice of quantum correlation measures. (ii) In the absence of an external applied magnetic field across the system under consideration, entanglement-separability monogamy scores will always be monogamous for the entire range of variable parameters considered here. In contrast, the information-theoretic monogamy scores can be non-monogamous for the same parametric considerations. (iii) With the increase in applied magnetic field, the monogamy scores with both kinds of quantum correlations exhibit an increase in the percentage of non-monogamous states. Also, this percentage on non-monogamous states with respect to information-theoretic kind of quantum correlations is  greater than the enetnalgment-separability kind.}
%
This is because all
the subsystem quantum correlations from information-theoretic kind will
always remain non-zero whereas several of the subsystem quantum
correlations from entanglement-separability kind will tend to zero for
increasing values of system parameters. (iv) We also observe that for higher
two-body interaction strengths the whole system and subsystem quantum
correlations for the spin system considered here will remain unaffected
by the external magnetic field and will depend only on three-body
interaction strengths.

In conclusion, it is customary to consider only two-body interactions
among the many-body quantum spin systems to study their relevance in
realizing several multiport quantum information protocols,
characterization of quantum correlations, detection of collective
signatures, properties of statistical features, etc. However, there also
exist many-body quantum spin systems with two or more spin interactions
in their Hamiltonian. We consider such a many-body quantum spin system
with two- and three-body interactions to study the shareability of
quantum information resources among its subsystems. Our study has a
plausible relevance to characterize the strengths of bipartite quantum
correlations and quantify the shareability among subsystems of a
multiparty quantum spin system which can lead to the realization of
networked multiport quantum information protocols.

\begin{acknowledgements}
RP and HSH acknowledge financial support from the Science and
Engineering Research Board (SERB), Government of India, under the
project grants CRG/2018/004811 and CRG/2021/008795. We acknowledge the
computations performed at \textit{AnantGanak}, the High Performance
Computing facility at IIT Dharwad.
\end{acknowledgements}

\appendix
\section{Quantum Correlation Measures}\label{sec:appendix}
In this appendix, we consolidate the definitions of (1) concurrence and
(2) logarithmic negativity chosen from entanglement-separability quantum
correlation measures and (3) quantum discord and (4) quantum work
deficit from information-theoretic kind which are used in this paper.  

\subsection{Concurrence}\label{App:concurrence} 
For an arbitrary two-particle state $\rho_{AB}$, concurrence is given
by~\cite{hill1997entanglement, wootters1998entanglement}

\begin{equation}
    C(\rho_{AB}) = \max \qty{0,\lambda_1 - \lambda_2 - \lambda_3 - \lambda_4},
    \label{eq:conc}
\end{equation} 
where the $\lambda_i$'s are the square roots of the eigenvalues of $\rho
\tilde{\rho}$ in the decreasing order. Here $ \tilde{\rho}$ is given by
$ \tilde{\rho} = (\sigma_y \otimes \sigma_y) \rho^* (\sigma_y \otimes
\sigma_y)$, where the complex conjugation is performed in the
computational basis, and $\sigma_y$ is the Pauli spin matrix in
$y$-direction. Concurrence vanishes for all separable states, whereas it
is maximum for maximally entangled states. For a pure bipartite state
$\psi_{AB}$, the concurrence is given by $2 \sqrt{\det \rho_A}$, where
$\rho_A$ is the reduced density matrix obtained by tracing out the
subsystem $B$ from $\psi_{AB}$~\cite{coffman2000distributed}.

\subsection{Logarithmic Negativity (LN)}\label{App:LN}

Logarithmic negativity~\cite{plenio2005logarithmic} is a computable
measure of entanglement defined as 
\begin{equation}
	\text{LN} (\rho_{AB}) = \log_2 \norm{\rho^{T_A}}_1,
  \label{eq:LN}
\end{equation}
where the partial transpose $T$ is taken over the subsystem $A$ and
$\norm{\cdot}$ is the trace norm. $\text{LN} (\rho_{AB})$ is $0$ for
separable states and $\text{LN} (\rho_{AB}) > 0$ for entangled states.
Logarithmic negativity can quantify the entanglement between two
partitions in a $N$-party quantum system. 

\subsection{Quantum Discord}\label{App:discord}
Quantum discord is a measure of quantum correlations beyond
entanglement~\cite{henderson2001classical, ollivier2001quantum}. It
utilizes the fact that the two equivalent definitions of classical
mutual information are not equal when their natural extensions are
considered within quantum theory.  The first definition of mutual
information of the quantum state $\rho_{AB}$, which defines total
correlation in the state, is given by 
\begin{equation}
    I (\rho_{AB}) = S(\rho_A) + S(\rho_B) - S(\rho_{AB}),
\end{equation}
where $\rho_A$ and $\rho_B$ are the reduced density matrices of
$\rho_{AB}$ and $S(\rho)$ represents the von Neumann entropy of the
state $\rho$. The second quantum version of mutual information, which
defines the classical correlation present in the system, is given by
\begin{equation}
    J(\rho_{AB}) = S(\rho_A) - S(\rho_{A|B}),
\end{equation}
where the second term is the {\em quantum conditional entropy}, and is
defined as 
\begin{equation}
    S(\rho_{A|B}) = \min_{\prod_i^B} \sum_i p_i S(\rho_{A|i})
\end{equation}
with the minimization being over all projection-valued measurements,
${\prod_i^B}$, performed on subsystem $B$. Here $p_i = {\rm tr}_{AB}
\qty[I_A \otimes \prod_i^B \rho_{AB} I_A \otimes \prod_i^B]$ is the
probability for obtaining the outcome $i$, and the corresponding
post-measurement state for the subsystem $A$ is $\rho_{A|i} =
\frac{1}{p_i} {\rm tr}_B \qty[I_A \otimes \prod_i^B \rho_{AB} I_A
\otimes \prod_i^B]$, where $I_A$ is the identity operator on the Hilbert
space of $A$.  The difference between the two quantities $I$ and $J$
quantifies the quantum correlation, which is defined as quantum discord
given by
\begin{equation}
    D(\rho_{AB}) = I(\rho_{AB}) - J(\rho_{AB}).
    \label{eq:disc}
\end{equation}
Note that for pure bipartite states, quantum discord reduces to the von
Neumann entropy of the reduced state of the smallest dimension among the
bipartition of the multiparty quantum system.

\subsection{Quantum Work Deficit}\label{App:work}
The amount of extractable pure states from a bipartite state
$\rho_{AB}$, under a set of global operations called the ``closed
operations'' (CO), is given by~\cite{oppenheim2002thermodynamical,
horodecki2003local, horodecki2005local, devetak2005distillation} 
\begin{equation} 
I_{\rm CO} = \log_2 \dim (\mathcal{H}) - S(\rho_{AB}), 
\end{equation} 
where the set of closed operations consists of unitary operations and
dephasing the bipartite state by a set of projectors, {$\prod_k$},
defined on the Hilbert space $\mathcal{H}$ of $\rho_{AB}$.  On the other
hand, considering the set of ``closed local operations and classical
communication'' (CLOCC), the amount of extractable pure states from
$\rho_{AB}$ is given by
\begin{equation}
    I_{\rm CLOCC} = \log_2 \dim (\mathcal{H}) - \min S(\rho'_{AB})
\end{equation}
Here, CLOCC consists of local unitary operations, dephasing by local
measurement on the subsystem, say, $B$, and communicating the dephased
subsystem to the other party, $A$, via a noiseless quantum channel. The
average quantum state, after the local projective measurement
{$\prod_k^B$} is performed on $B$, can be written as $\rho'_{AB} =
\sum_k p_k \rho_{AB}^k$ with $\rho_{AB}^k$ and $p_k$. The minimization
in $I_{\rm CLOCC}$ is achieved over all the complete sets {$\prod_k^B$}.
The ``one-way'' quantum work deficit is then defined as
\begin{align}
    W^{\leftarrow} (\rho_{AB}) &= I_{\rm CO} - I_{\rm CLOCC} \nonumber \\ 
                               &= \min_{\qty{\prod_k^B}}
                               \qty[S(\rho'_{AB}) - S(\rho_{AB})].
    \label{eq:wd}
\end{align}
Here, the arrow in the superscript starts from the subsystem over which
the measurement is performed. Note that for pure bipartite states,
quantum work deficit reduces to the von Neumann entropy of the reduced
state of the smallest dimension.

\bibliography{references}
\bibliographystyle{apsrev4-2}

\end{document}